\documentclass[12pt, a4paper]{article}

\usepackage{amsmath} \usepackage{amsthm} \usepackage{amsfonts}
\usepackage{amssymb} \usepackage{graphicx, rotating} \usepackage{epstopdf}
\usepackage{epsfig} \usepackage{latexsym} \usepackage{graphicx}
\usepackage{color} \usepackage{amsmath,amssymb} \usepackage{cite}
\usepackage{slashed} \usepackage{float}
\numberwithin{equation}{section}
\usepackage{pgfplots}
\pgfplotsset{compat=1.9}
\usetikzlibrary{ calc, decorations.pathmorphing, decorations.markings }

\usepackage{placeins}
\usepackage{standalone} 
\usepackage{physics} 
\usepackage{textgreek}
\definecolor{bluscuro}{rgb}{0.15, 0.2, .85}
\definecolor{bluchiaro}{cmyk}{1,.3,0.,0.1}
\definecolor{colorblind1}{HTML}{648FFF}
\definecolor{colorblind2}{HTML}{785EF0}
\definecolor{colorblind3}{HTML}{DC267F}
\definecolor{colorblind4}{HTML}{FE6100}
\definecolor{colorblind5}{HTML}{FFB000}
\usepackage[colorlinks=True, citecolor=bluscuro, linkcolor=black,urlcolor=bluscuro,linktocpage]{hyperref}

\newcommand{\eq}[1]{Eq.~(\ref{#1})}
\newcommand{\nn}{\nonumber}
\newcommand{\PI}{\text{\textpi}}
\newcommand{\iu}{\mathrm{i}\mkern1mu}

 \newcommand{\bs}{s}
 \newcommand{\os}{{ s^\prime}} 
\newcommand{\Mh}{{\cal M}}

\newcommand{\doa}{\hspace{-3pt}\footnotesize\begin{array}{c}\smallfrown\\[-7pt]\smallsmile\end{array}\hspace{-3pt}}

\newcommand{\stl}{strictly tree-level}

  \def\I{\text{i}} \def\dd{\text{d}}

\usepackage[utf8]{inputenc} \usepackage[utf8]{inputenc}
\usepackage[english]{babel} \usepackage{fancyhdr} \usepackage{amsmath}
\usepackage{amssymb} \usepackage{amsfonts} \usepackage{psfrag}
\usepackage{graphicx,wrapfig} \usepackage[bf,footnotesize]{caption2}
\usepackage{cite} \usepackage{bm}
\usepackage{comment}

\usepackage[sort&compress,numbers, merge]{natbib} \usepackage{amsthm,bm}
\usepackage{gensymb}

\usepackage{physics} \usepackage{array} \usepackage{tcolorbox, mathtools}
\usepackage{soul} \usepackage{feynmp-auto}

\title{Non-Forward UV/IR Relations} 
\begin{document}

\vspace*{-2cm} \begin{flushright} \vspace*{2mm} \today \end{flushright}

\begin{center} \vspace*{15mm}

\vspace{1cm} {\large \bf  Non-Forward UV/IR Relations} \\
\vspace{1.4cm}

{ Carl Beadle$\,^{a}$, Giulia Isabella$\,^{a,b}$, Davide Perrone$\,^{a}$, Sara Ricossa$\,^{a}$, Francesco~Riva$\,^{a}$, Francesco Serra$\,^{a,c,d}$}

 \vspace*{.5cm} {\it
$^a$ D\'epartment de Physique Th\'eorique, Universit\'e de Gen\`eve\\
$^b$ Mani L. Bhaumik Institute for Theoretical Physics, Department of Physics and Astronomy, University of California Los Angeles, Los Angeles, CA 90095, USA\\
$^c$ Scuola Normale Superiore, Pisa\\
$^d$ Department of Physics and Astronomy, Johns Hopkins University}

\vspace*{10mm} 
\begin{abstract}\noindent\normalsize
We study bounds  arising from the analyticity and unitarity of scattering amplitudes in the context of  effective field theories with massless particles. 
 We provide an approach that only uses dispersion relations away from the forward limit. 
 This is suitable to derive constraints in the presence of gravity, in a way that is robust with respect to radiative corrections. 
 Our method not only allows us to avoid the Coulomb pole, but also the singularities associated with calculable loop effects, which would otherwise diverge.

\end{abstract}

\vspace*{3mm}
\end{center}
\newpage 

\tableofcontents

\newpage

\section{Motivation}

Which low-energy  phenomena can be described consistently within quantum field theory?
This question plays an important role in  contexts such as  physics beyond the standard model or quantum gravity. There, our comprehension of Nature is formulated in terms of effective field theories~(EFTs), where the leading order  terms are fixed by past experiments,  while higher order  terms provide experimentally testable signatures for the future, as they reveal information about  UV-completions.

Within the landscape of such possible signatures, understanding which ones are allowed in quantum field theory (QFT) is an important goal of phenomenology.
In practice, this translates into discerning between theories which admit a UV-completion, and those that do~not.
While it is impossible to  explore  the multitude  of explicit UV $\to$ IR maps, the broadest QFT hypotheses in the UV already contain enough information to vastly constrain the space of allowed EFTs.
Exploiting  unitarity and the analytic properties of scattering amplitudes---implied by the Lorentz-invariant causal structure of QFT~\cite{Bros:1964iho,Eden:1971fm}---it is possible to characterize the space of UV-completable EFTs, via positivity bounds and dispersion relations on their Wilson coefficients~\cite{Adams:2006sv,Arkani-Hamed:2020blm,deRham:2017avq,Bellazzini:2020cot,Sinha:2020win,Tolley:2020gtv,Caron-Huot:2020cmc,EliasMiro:2022xaa,Chowdhury:2021ynh}.
Applications of these techniques have been found in QCD~\cite{Pham:1985cr,Ananthanarayan:1994hf,Pennington:1994kc,Bijnens:1997vq,Ananthanarayan:2000ht,Colangelo:2001df,Manohar:2008tc,Mateu:2008gv,Guerrieri:2019rwp,Guerrieri:2020bto,Zahed:2021fkp,Figueroa:2022onw}, quantum gravity~\cite{Camanho:2014apa,Bellazzini:2015cra,Hamada:2018dde,Bonifacio:2018vzv,Bellazzini:2019xts,Melville:2019wyy,Tokuda:2020mlf,Herrero-Valea:2020wxz,Edelstein:2021jyu,Bern:2021ppb,Bellazzini:2021shn,Arkani-Hamed:2021ajd,Caron-Huot:2022ugt,Serra:2022pzl,Noumi:2022wwf,Herrero-Valea:2022lfd,Chiang:2022jep,Bellazzini:2022wzv,Caron-Huot:2021rmr,Caron-Huot:2022jli,Henriksson:2022oeu,Albert:2022oes,Fernandez:2022kzi,Albert:2023jtd,Ma:2023vgc,Arkani-Hamed:2021ajd,Eckner:2024ggx,Haring:2023zwu} and physics beyond the Standard Model~\cite{Bellazzini:2018paj,Zhang:2018shp,Remmen:2020vts,Karateev:2022jdb,Haring:2022sdp,Hebbar:2020ukp,Bonnefoy:2020yee,Riembau:2022yse,Acanfora:2023axz,Bertucci:2024qzt,EliasMiro:2023fqi,McCullough:2023szd}.
These tools have been used to prove the a-theorem in 4-dimensions~\cite{Komargodski:2011vj,Hong:2023zgm} and to show that EFTs  dominated by soft interactions~\cite{Camanho:2014apa,Englert:2019zmt,Bellazzini:2020cot}, such as
EFTs with weakly broken Galilean symmetry~\cite{Tolley:2020gtv,Caron-Huot:2020cmc},  or higher-spin particles  parametrically lighter than the EFT cutoff~\cite{Bellazzini:2019bzh,Bellazzini:2023nqj,Bertucci:2024qzt}, are inconsistent.

Here we focus on EFT positivity bounds, as derived by dispersion relations for the $2\to2$ scattering amplitude for spin-0 particles, in the Mandelstam variable $s$ and for fixed $t$.  
The bounds stem from writing this quantity in an IR representation---where it can be computed within the EFT as a function of the Wilson coefficients---and a UV representation, where unitarity implies positivity.  
In this context, several approaches have been proposed to extract bounds, by including information about the finite $t$ behaviour. 
The simplest method involves a Taylor-expansion  at $t=0$~\cite{Arkani-Hamed:2020blm,Caron-Huot:2020cmc,deRham:2017avq}, and as such, is only applicable when the amplitude is analytic in~$t=0$.
This requirement fails to be satisfied in many physically relevant theories that have  massless particles included in the spectrum. For instance, in the presence of gravity, beside the Coulomb pole $\delta\mathcal{M}\propto s^2/t$, computable-loop effects involving gravitons lead to contributions to the amplitude at small $t$ of the form~\cite{gravityloops},
\begin{equation}\label{eq:singularities0}
    \delta\mathcal{M}(s,t)\propto  s^2\log(-s)   \left( - t \right)^{\frac{d -
4}{2}} \quad (\times \log(-t)\,\,\, \textrm{
in even dimensions}).
\end{equation}
These effects are important because, due to their non-analyticity in $s$, they appear in \emph{all} dispersion relations. Then, the non-analyticity in $t$ prevents dispersion relations, or their derivatives, from approaching the forward limit $t=0$.

In Ref.~\cite{Caron-Huot:2021rmr,Caron-Huot:2022ugt} this issue was in part overcome,  by treating dispersion relations not as functions of $t$, but as distributions to be smeared against appropriate functional measures. The method proposed there is designed to avoid the Coulomb singularity, but not the loop effects of \eq{eq:singularities0}, which are still evaluated at $t=0$ and are technically infinite. In this  sense,  perturbation theory fails in the context of positivity bounds.

In this work, we describe under what circumstances smeared distributions are sensitive to the individual coefficients of a Taylor expansion, such as the amplitude stemming from an~EFT.
We then propose a procedure to obtain positivity bounds, which  \emph{completely avoids the forward limit}. 
These can be consistently used to derive bounds in any theory (including gravity) beyond the \stl~case. 
Our technique is rooted in the spirit of EFTs: we include operators of sufficiently large dimension and work at sufficiently high loop order to allow us to obtain results of any desired accuracy.

This work provides a proof of principle showing that the
forward limit can be completely avoided, in order to extract meaningful positivity bounds.
Away from the forward limit, loop effects are finite  and can be treated perturbatively.

\vspace{0.5cm}

In Section~\ref{sec:dispersion} we introduce dispersion relations and review positivity constraints in the forward limit. In Section~\ref{sec:finitetimprovement} we discuss smeared dispersion relations in light of the M\"untz-Sz\'asz theorem, which encapsulates the differences between smearing and Taylor expansion. We then
present an  algorithm to mediate between the two, based solely on finite $t$ relations.
The numerical results are collected in Section~\ref{sec:boundswc}. We conclude and discuss further avenues in Section~\ref{sec:conclusion}.
We use the appendices to go into more details on the M\"untz-Sz\'asz theorem~\ref{app:MS}, on designing our algorithm~\ref{app:MF} and on the numerical procedure to extract bounds~\ref{app:smerding}. Finally, in appendix~\ref{sec:apppol} we discuss alternative possibilities for the smeared dispersion relations.

\section{Dispersion relations}\label{sec:dispersion}

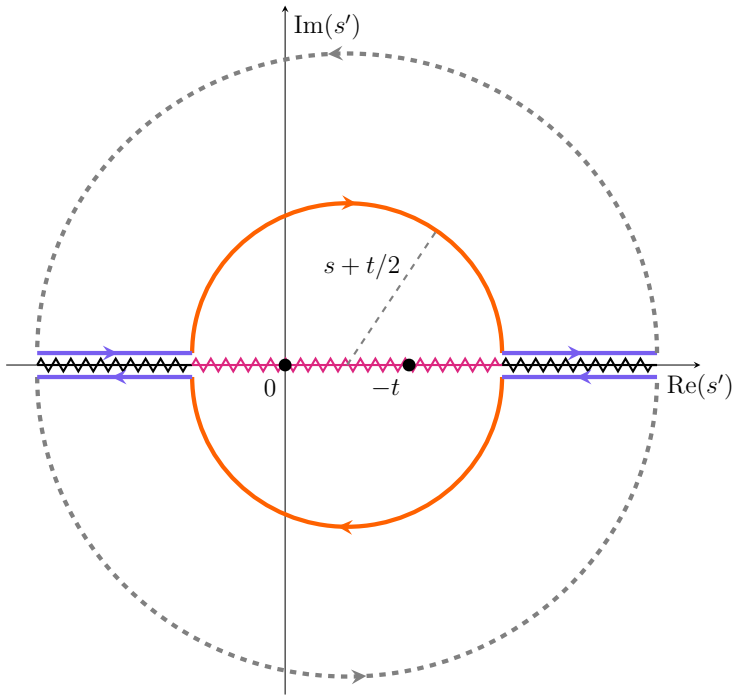
\begin{figure}[h]
    \centering
    \begin{tikzpicture}[scale=0.8]

        \begin{axis}[
            width=13cm,
            height=13cm,
            axis lines = center,
            ticks=none,
            xlabel=Re$(s')$,
            ylabel=Im$(s')$,
            xlabel style={below},
            ymin=-5.5,
            ymax=+6,
            xmin=-4.5,
            xmax=+6.7
        ]          
            \addplot [black, mark = *, line width=2pt] coordinates {( 0, 0)} {};
            \addplot [black, mark = *, line width=2pt] coordinates {( 2, 0)} {};

            \node [below left, black] at (axis cs:  0, -0.1) {$0$};
            \node [below left, black] at (axis cs:  2, -0.1) {$-t$};
            \node [below left, black] at (axis cs:  2, 2) {$s+t/2$};
        
            \addplot [
                line width=1pt,
                color=black,
                postaction={decorate},
                decoration={zigzag, amplitude=3pt, segment length=7pt}
            ] coordinates { (-4,0) (-1.5,0) };
            \addplot [
                line width=1pt,
                color=black,
                postaction={decorate},
                decoration={zigzag, amplitude=3pt, segment length=7pt}
            ] coordinates { (3.5,0) (6,0) };
            \addplot [
                line width=1pt,
                color=colorblind3,
                postaction={decorate},
                decoration={zigzag, amplitude=3pt, segment length=7pt}
            ] coordinates { (-1.5,0) (3.5,0) };
            \draw [
                line width=2pt,
                color=colorblind4,
                postaction={decorate},
                decoration={markings, mark=at position 0.52 with {\arrow{stealth}}}
            ] (axis cs: -1.5, 0.2) arc (180:0:250) -- (axis cs: 3.5, 0.2);
            \addplot [
                line width=2pt,
                color=colorblind2,
                postaction={decorate},
                decoration={markings, mark=at position 0.52 with {\arrow{stealth}}}
            ] coordinates { (-4,0.2) (-1.5,0.2) };
           \addplot [
                line width=2pt,
                color=colorblind2,
                postaction={decorate},
                decoration={markings, mark=at position 0.52 with {\arrow{stealth}}}
            ] coordinates { (3.5,0.2) (6,0.2) };
            \draw [
                dashed,
                line width=2pt,
                color=gray,
                postaction={decorate},
                decoration={markings, mark=at position 0.52 with {\arrow{stealth}}}
            ] (axis cs: 6, 0.2) arc (0:180:500) -- (axis cs: -4, 0.2);
            \draw [
                dashed,
                line width=2pt,
                color=gray,
                postaction={decorate},
                decoration={markings, mark=at position 0.52 with {\arrow{stealth}}}
            ] (axis cs: -4, -0.2) arc (-180:0:500) -- (axis cs: 6, -0.2);
            \draw [
                line width=2pt,
                color=colorblind4,
                postaction={decorate},
                decoration={markings, mark=at position 0.52 with {\arrow{stealth}}}
            ] (axis cs: 3.5, -0.2) arc (0:-180:250) -- (axis cs: -1.5, -0.2);
            \addplot [
                line width=2pt,
                color=colorblind2,
                postaction={decorate},
                decoration={markings, mark=at position 0.52 with {\arrow{stealth}}}
            ] coordinates { (-1.5,-0.2) (-4,-0.2) };
           \addplot [
                line width=2pt,
                color=colorblind2,
                postaction={decorate},
                decoration={markings, mark=at position 0.52 with {\arrow{stealth}}}
            ] coordinates { (6,-0.2) (3.5,-0.2) }; 
            \addplot [
                dashed,
                line width=1pt,
                color=gray
            ] coordinates { (1,0) (2.46,2.27) };
        \end{axis}
    \end{tikzpicture}
    \caption{\it \footnotesize The $2\rightarrow2$ scattering amplitude is analytic in the upper (and by crossing, lower) half plane in $s'$.
    The contour in red is the IR arc~\eq{eq:archdeft} that runs within the EFT validity region, whereas the contour in blue denotes the arc in the UV~\eq{eq:archdeftUV}. In magenta the IR part of the branch cut, due to loops in the EFT, in black its UV part, associated also to particle exchange in the unknown UV theory.
    We indicate the two subtraction points at $s'=0$ and $s'=-t$.}\label{fig:analyticAmplitude}
\end{figure}

We are interested in the $2\to2$ scattering amplitude for spin-0 massless particles, i.e. Goldstone bosons, and study this amplitude in the complex Mandelstam variable $s\in \mathbb{C}$ plane, for fixed~$t<0$. The Lorentz-invariant causal structure implies that the only non-analyticities are confined to the real $s$ axis and are associated with physical phenomena~\cite{Bros:1964iho} (see Figure~\ref{fig:analyticAmplitude}). 
We define arcs in their \emph{IR representation} as contour integrals in $s$,
\begin{equation}\label{eq:archdeft} 
    a_n(\bs,t) \equiv
    \int_{\doa}\frac{\dd\os}{2\pi \iu \, \os}\frac{\Mh(\os,t)}{[\os \, (\os+t)]^{n+1}} \,,\quad
    n\geq0\,.
\end{equation}
As shown in Figure~\ref{fig:analyticAmplitude}, the contour of integration `$\doa$' is given by the union of two semi-circles of radius $|\bs +t/2|$, centered at~$-t/2$. Notice that the measure and subtraction is manifestly $s$-$u$ crossing symmetric.
Arcs probe the theory at finite energy, and are particularly  appropriate in the presence of massless particles, where the branch cut extends all the way to the origin in $s$~\cite{Luty:2012ww,Bellazzini:2020cot}. An assumption that is often made in the literature is that below the mass-gap of the EFT $s\leq M^2$, the theory is weakly coupled and  the discontinuity along this branch cut is  negligible. In this situation, arcs would reduce to the residues of the $n$-subtracted amplitude at $s=0$ and $s=-t$. 
In this article we develop a tool that is suitable to deal with the IR effects, and therefore we keep the arc as defined above, although in practice we will often work with the tree-level formulae as well.

Exploiting the analyticity properties of the amplitude in the complex $s$-plane, the integration contour $\doa$ can be deformed into another contour that encompasses the discontinuities on the positive $\os\geq \bs$ and negative $\os\leq -\bs-t $ real axis, as well as two semicircles at infinity enclosing the whole upper and lower-half planes (see Figure~\ref{fig:analyticAmplitude}).
We further assume that the latter contributions vanish, as implied by generalisations of the Froissart-Martin bound~\cite{Froissart:1961ux,Martin:1962rt,Jin:1964zza} to theories involving gravity~\cite{Caron-Huot:2021rmr,Haring:2022cyf}, corresponding to  the asymptotic behavior
\begin{equation}
    \lim_{|s|\to \infty} \Mh(s,t)/s^2 =0\,.
\end{equation}

Moreover, the property of $s$-$u$ crossing symmetry of both the amplitude 
$\Mh(\bs,t)=  \Mh(-\bs-t,t)$ 
and the measure in \eq{eq:archdeft}, combined with real analyticity
$\Mh(\bs,t)=  \Mh^*(\bs^*,t)$, allows us to relate the integrals along the positive and negative real axis, obtaining,
\begin{equation}\label{eq:archdeftUV} 
    a_n(\bs,t) =\frac{1}{\pi}\int_{\bs}^\infty \frac{\dd\os}{\os} \frac{\left(2\os+t\right)}{[\os(\os+t)]^{n+1}}\,\textrm{Im} \,\Mh(\os,t)\,.
\end{equation} 
At this point, unitarity in the forward $t=0$ limit directly translates into $\textrm{Im} \,\Mh(\os,0)\geq 0$, and the condition $a_n(s,0)\geq0$. 
As anticipated, for theories including massless particles, the forward limit may be divergent. To exploit unitarity away from the forward limit, we project the amplitude onto partial waves,
\begin{gather}\label{eq:pwexp} 
    \Mh(\bs,t)= \sum_{\ell=0}^\infty \mathcal{M}_\ell(\bs) \, {\cal
    P}_\ell\Big(1+\frac{2 \, t}{\bs}\,\Big) \,,
\end{gather} 
where 
${\cal P}_\ell(\cos\theta)=\,_2F_1\left(-\ell, \ell + d - 3, (d - 2)/2, (1 -\cos\theta)/2\right)$
are the Gegenbauer polynomials in $d$ space-time dimensions, which reduce to Legendre polynomials in $d=4$. The appeal of using this basis is that S-matrix unitarity is diagonal in the partial waves, and  directly implies $\text{Im}\,\mathcal{M}_\ell(\bs)>0$.
In \eq{eq:pwexp} we absorb any normalisation factors into the definition of $\Mh_\ell(s)$.
Note that, for identical scalars, the amplitude is symmetric under $\cos \theta\to - \cos\theta$ and the partial waves expansion has support only on even $\ell$'s.
By projecting the right hand side of \eq{eq:pwexp} onto partial waves, we are also expressing the arcs in a \emph{UV representation}

\begin{eqnarray}\label{eq:archdeftUV2} 
    a_n(\bs,t) &=&\frac{1}{\pi}\int_{\bs}^\infty\frac{\dd\os}{\os} \, \sum_{\ell=0}^\infty  \rho_\ell(\os) \, \frac{\left(2\, \os+t\right)}{[\os(\os+t)]^{n+1}} \, {\cal P}_\ell\Big(1+\frac{2\, t}{\os}\,\Big)\,\nn \\ &\equiv&\left\langle \frac{\left(2 \, \os+t\right)}{[\os(\os+t)]^{n+1}}\, {\cal P}_\ell\Big(1+\frac{2\, t}{\os}\,\Big) \right\rangle \,, 
\end{eqnarray} 
where we have defined the average 
\begin{eqnarray}\label{eq:average}
    \bigg\langle \cdots\bigg\rangle=\frac{1}{\pi}\int_s^\infty\frac{\dd\os}{\os} \, \sum_{\ell=0}^\infty \rho_\ell(\os)  \, ( \,\cdots)\,,
\end{eqnarray}
and $\rho_\ell(s)\equiv\textrm{Im}\,\Mh_\ell>0$ is the spectral density, which provides a positive measure for the average $\langle \, \cdots \rangle$.  
This  positivity implies constraints for the UV representation of arcs of \eq{eq:archdeftUV2}, which are inherited by the IR representation (\ref{eq:archdeft}), implying bounded coefficients within the~EFT.

\subsection{IR arcs}\label{subsec:TreeLevel}

In this section, we explicitly compute the arcs in their IR representations, for exact (massless) Goldstone bosons interacting among themselves and gravitationally in a theory with a mass gap $M$.
At sufficiently low energy $E\ll M$ the theory is weakly coupled and well described by an effective  Lagrangian with interactions ordered by a derivative expansion. 
We organise the  amplitude  as,
\begin{equation}
    \Mh=\Mh^{\text{tree}}_{\text{EFT}}+\Mh^{\text{loops}}\,.
\end{equation}
The tree-level contribution includes graviton exchange and higher order contact interactions,
\begin{equation}\label{eq:amptreeXY}
    \Mh^{\text{tree}}_{\text{EFT}} = \kappa^2 \left( \frac{u \, t}{s} + \frac{s \, u}{t} + \frac{s \, t}{u} \right)+\sum_{p\geq 1}\sum_{q=0}^p g_{2p+q,q} \left(\frac{s^2 + t^2 + u^2}{2}\right)^{p-q}\left(s  t u\right)^{q} \,,
\end{equation}
with $ \kappa^2 \equiv \frac{1}{M_{\text{P}}^{d-2}} $ denoting the gravitational constant in $d$-dimensions, where $M_{\text{P}}$ is the Planck scale $M_\text{P}^{-2}=8\pi G$ in $d=4$ dimensions.\footnote{
A constant term in the amplitude \eq{eq:amptreeXY}, as well as a pole associated with the scalar exchange  are forbidden by the Goldstone Boson shift symmetry. 
Furthermore, the contribution from gravity is limited to the amplitude emerging from minimal coupling because the 3-point interaction between two scalars and one graviton is unique.
} 
The coefficients $g_{n,q}$ scale as $1/M^{2n}$ in units of the mass gap.
We will refer to an amplitude with $\Mh^{\text{loops}}=0$ (which is equivalent to the limit $\kappa, g_{n,q}\to 0$, with finite ratios) as ``\stl''.

\paragraph{Strictly tree-level.}
The tree-level part of the amplitude is analytic in both $s$ and $t$ away from the origin and the arcs reduce to  the sum of residues at $s=0$ and $s=-t$. 
From \eq{eq:amptreeXY} we can then find an all-order expression for the arcs, 
\begin{align}\label{eq:generalarcs} 
    a^{\text{tree}}_n(s,t) &= (\textrm{Res} \, |_{\os=0}+\textrm{Res} \, |_{\os=-t})\frac{\Mh(\os,t)}{\os[\os(\os+t)]^{n+1}}\nn \\
    &= -\frac{\kappa^2}{t} \, \delta_{n,0} +\sum_{p=1}^\infty\sum_{q=0}^pg_{2p+q,q}(-t)^{2(p-n-1)+q} \, \binom{p-q}{n+1-q}.
\end{align} 
Notice that the gravity pole only appears in the zeroth arc,
\begin{eqnarray}\label{eq:arcsexplicit}
    a^{\text{tree}}_0(\bs,t)&=&-\frac{\kappa^2}{t}+\sum_{n=1}^\infty[ \, n\, t^{2n-2} \, g_{2n,0}-
    t^{2n-1} \, g_{2n+1,1}] \nn \\ 
    &=&-\frac{\kappa^2}{t} + g_{2,0} - g_{3,1}t+2 \, g_{4,0}t^2 + \cdots \,, 
\end{eqnarray} 
while all of the other arcs are polynomials in $t$, for example:
\begin{equation}\label{eq:arc12tree}
\begin{split}
    a^{\text{tree}}_1(\bs,t) &= g_{4,0}-t g_{5,1}+3
    t^2 g_{6,0}+t^2 g_{6,2}+\cdots  \\
    a^{\text{tree}}_2(\bs,t) &= g_{6,0}-t g_{7,1}+4 t^2
    g_{8,0}+t^2 g_{8,2}+\cdots \,
\end{split}
\end{equation}

Without gravity and in the \stl~limit, these IR arcs are all polynomials in~$t$.
Then one can expand both the IR and UV arcs  in $t$ using a Taylor series, which would provide a set of equations for the EFT coefficients in terms of UV averages \eq{eq:average}. 
For instance, from \eq{eq:archdeftUV} and \eq{eq:generalarcs} at $t=0$ we may read
\begin{equation}
    g_{2,0}=\left<\frac{1}{\os}\right>\,,\; \;
    g_{4,0}=\left<\frac{1}{\os^3}\right>\, \;\; \text{and} \;\;
    g_{6,0}=\left<\frac{1}{\os^5}\right>\,. 
\end{equation} 
Since $\os\geq s$ and the measure in the average is positive, the Wilson coefficients surviving in the forward limit must monotonically decrease in units of $s$, and must also satisfy two-sided bounds $g_{2,0}\geq g_{4,0}s^2\geq g_{6,0}s^4\geq 0$~\cite{Bellazzini:2020cot}.

Moreover, there is a redundancy in these equations because many EFT coefficients $g_{n,q}$ appear in multiple arcs, multiplying different powers of $t$.
Indeed, from \eq{eq:arcsexplicit} and \eq{eq:arc12tree}, we observe that $g_{2,0}$ and $g_{3,1}$ appear only in $a_0$, while $g_{4,0}$ and $g_{5,1}$ also appear in $a_1$. 
On the other hand the coefficient $g_{6,2}$ appears uniquely in $a_1$, while there are none in $a_2$. 
This redundancy implies the existence of ``{null constraints}'': non-trivial relations between UV representations of different arcs.
For instance, from \eq{eq:arcsexplicit} (with $\kappa=0$) and \eq{eq:arc12tree}, we see that for the IR arcs,
\begin{equation}
 \left.(\partial_t^2 a^{\text{tree}}_0  -4 a_1^{\text{tree}})\right|_{t=0}=0\,.
\end{equation}
These relations can be thought as constraints on the UV measure $\rho_\ell (s)$ appearing in the average. Using them, one finds two-sided bounds also for the coefficients that vanish in the forward limit, such as $g_{3,1}$~\cite{Tolley:2020gtv,Caron-Huot:2020cmc}.

\vspace{5mm}

With gravity, the first arc \eq{eq:arcsexplicit} is singular at $t=0$, so the approach based on the Taylor expansion fails -- more on this in the next section.
On the other hand, gravity at tree level does not enter in any other arc $a_n$ with $n\geq1$, all of which are still polynomials. Therefore, in the \stl~limit, it is still possible to build null constraints using these higher arcs.

\paragraph{Loop-level.}

The 1-loop effects in $\Mh^{\text{loops}}$ are known, see e.g.~\cite{Bellazzini:2021oaj,Caron-Huot:2024tsk,Arkani-Hamed:2021ajd,gravityloops} in the context of positivity.
Since, by construction, the EFT is weakly coupled in the far IR, these effects have small coefficients, of the order $g^2/16\pi^2$ or $\kappa^4/16\pi^2$ in the case of gravity. For this reason, they are mostly neglected in the EFT positivity literature, where one works in the \stl~limit.

Despite the small coefficients, loop effects qualitatively modify the analytic structure of amplitudes, and  therefore play an important role in dispersion relations, see \eq{eq:singularities0}.
For instance, at order $\kappa^4$ in $d=5$ dimensions---where the computation is particularly well-defined---the amplitude develops features such as,
\begin{eqnarray}\label{eq:blabla}
\Mh^{\text{loops}}\propto \sqrt{-t}\,s^2 \log s\,.    
\end{eqnarray}
Because of the non analyticity in $s$, such effects appear in \emph{all} arcs. Then, the non-analyticity in $t$ prevents the dispersion relations from being Taylor expanded. This implies that also all null constraints, obtained using arcs $a_n$ with $n\geq 1$, are singular.

This fact forces us to  rethink both dispersion relations and null constraints, while entirely avoiding the $t=0$ limit.

\section{Positivity at finite $t$}\label{sec:finitetimprovement}

When equating the IR and UV representations of the arcs (\eq{eq:archdeft} and \eq{eq:archdeftUV2}), it is 
important to understand in what sense the partial wave expansion correctly reproduces the amplitude.
Pointwise convergence (convergence at each point in $t$) is not always guaranteed.
Indeed, in the presence of gravity or loop effects, it is of course impossible to reproduce non-analytic structures, such as the $1/t$ pole or branch cuts, using only polynomials.
Instead, the series might converge in a weaker sense, meaning that given  appropriate measures $\dd\mu(t)\equiv f(t)\dd t$, the integrals,
\begin{equation}\label{eq:projection} 
    \int_{-t_{\text{max}}}^0 \dd\mu(t) \, a_n(\bs,t),
\end{equation}
with $0<t_{\rm max}<1$, will converge both in the IR and UV representations. The arcs $a_n(s,t)$ are therefore treated as distributions.
Hence, rather than Taylor expanding dispersion relations and comparing different powers of $t$ (as illustrated above), we shall project (smear) dispersion relations against an appropriate basis of  functions $f_i(t)$. 
The projected dispersion relations shall shape the allowed space of EFT parameters~\cite{Caron-Huot:2021rmr}. 

\subsection{Improving arcs to circumvent the M\"untz-Sz\'asz theorem}\label{subsec:MS}

There is a crucial difference between the Taylor expansion method and using the smeared relations that we should first discuss.
Indeed, the Taylor expansion is in one-to-one correspondence with the EFT approach, and the $t$-expanded arcs directly imply conditions on the EFT coefficients. Does a suitable basis of functions $f_i(t)$ exist, such that the smeared IR arcs  are also projected directly onto the EFT coefficients? 

To answer this question we rely on the fact that the space of continuous functions in $t\in[-t_{\text{max}},0]$ constitutes a vector space, albeit an infinite dimensional one. 
While many of the properties of finite dimensional vector spaces extend to infinite dimensional ones, the M\"untz-Sz\'asz theorem~\cite{Muntz1914,sasz} represents a striking difference between the two cases.
The monomials $\{1,t,t^2,t^3,\cdots\}$ are a basis of continuous functions on an interval. 
The M\"untz-Sz\'asz theorem states that the same is true for the generic basis,
\begin{equation}
    \{t^{\lambda_0},t^{\lambda_1},t^{\lambda_2},\cdots\} 
\end{equation}
with parameters $0=\lambda_0<\lambda_1 <\cdots $, iff
\begin{equation}
    \quad \sum_{i=1} \frac{1}{\lambda_i}\to \infty \,.
\end{equation}
In other words, this theorem quantifies how many ``missing terms'' a basis of monomials is allowed to have to still approximate a function or distribution arbitrarily well.  
For instance, it implies that any basis missing a few monomials, such as the set $\{1,t,t^3,t^4,t^5,\cdots\}$ that does not contain $t^2$, is still a basis. 

In  a Taylor expansion it makes sense to expand two equal expressions (such as the UV and IR representations of the arcs) in powers of $t$ and then compare their coefficients. 
Instead, when we talk about distributions under a measure, there is  no absolute meaning associated with the coefficient of a particular power of $t$, because it can just be re-written in terms of the other powers.
For instance a function $g_{2,0}-g_{3,1}t+2 g_{4,0}t^2+\cdots$ like the one appearing in~$a_0$, may as well be re-written as $\tilde g_{2,0}+2 \tilde g_{4,0}t^2+\cdots$, without the linear term in $t$, but instead with some new coefficients $\tilde g_{n,q}$. 
Therefore, \emph{without any extra assumptions or information}, it is not possible to extract unique coefficients of an infinite  Taylor series from an integrated distribution.

To understand what type of information might be needed, it is interesting to see how the M\"untz-Sz\'asz theorem works in practice. 
For instance, in the above example, how large would  the coefficients $\tilde{g}_{n,q}$ have to be relative to $g_{n,q}$ in order to be able to accurately describe the same function?
In Appendix~\ref{app:MS}, we discuss this question in detail and provide some extra examples. 
We find that the coefficients $\tilde g_{n,q}$ must grow exponentially fast in $n$  w.r.t.\ the original coefficients, see also Ref.~\cite{Trefethen:2023:SLE}. 
This means that mild assumptions about the asymptotic behaviour of the coefficients is enough to prevent them from being expressible in terms of higher ones.

For instance, the underlying assumption behind every EFT---that after a certain order the coefficients start becoming smaller in units of the energy---would be enough to limit the impact of the M\"untz-Sz\'asz theorem. 
On the other hand, without making use of such an assumption, one might wonder if it even makes sense to talk about bounds on the coefficients of the EFT from smearing, as the tree-level approximation is identical to a Taylor expansion. 

Even without this assumption, in the case of dispersion relations, extra information \emph{is} available. Full crossing symmetry implies that the same EFT coefficients appear in different dispersion relations.
The easiest way to see how this information can be used, is to algorithmically use dispersion relations themselves to subtract off all of the higher coefficients. 
It is indeed possible to find coefficients $c_{n,k}$ to build an \emph{improved arc},
\begin{equation}\label{eq:geralimp} 
    a_0^\text{imp}(s,t) =
    \sum_{n,k}c_{n,k}\frac{t^{2n+k}}{k!}\partial_t^ka_n(s,t)\,, 
\end{equation} such
that in its IR representation, only a finite number of terms appear. 
For instance, in the \stl~limit, the first arc can be improved to
\begin{equation}\label{eq:arcimpIR} 
    a_0^\text{imp}(s,t) =-\frac{\kappa^2}{t}+
    g_{2,0}-g_{3,1}t\,,
\end{equation}
cleared of any higher order term. 
The improved arc lacks the freedom to reproduce the monomial $t$ with other coefficients and the M\"untz-Sz\'asz theorem, as used above, does not apply. 
The resulting dispersion relations can be projected on test functions to unambiguously extract bounds on $g_{2,0}$ and $g_{3,1}$ in terms of $\kappa$. 
Similarly, an improved arc,
\begin{equation}\label{eq:arcimpIR2} 
    a_0^\text{imp}(s,t) =-\frac{\kappa^2}{t}+
    g_{2,0}-g_{3,1}t+2g_{4,0}t^2\,,
\end{equation}
allows one to include also $g_{4,0}$ in the projected dispersion relations, and so on.

The idea of improved arcs was already introduced in Ref.~\cite{Caron-Huot:2021rmr}, where an improved $a_0$ arc of the form \eq{eq:arcimpIR} was achieved via the linear combination,
\begin{equation}\label{eq:CHimp} 
    a_0(s,t) - \sum_{n\geq 1}t^{2n} \, \left(n \, a_n(s,0)- t \,  \partial_t a_n(s,t) \, \Big|_{t=0} \, \right). 
\end{equation}
Crucially, this expression  uses higher arcs and their first  derivatives at $t=0$. 

Loop effects introduce corrections to this formula when evaluated on the IR side.
In particular, all couplings $g_{n,q}$ re-appear in a non-linear way, at all orders in the perturbative expansion.
They now multiply each other so that their effect cannot be removed by a simple algebraic operation on arcs. 
In principle, one could take a perturbative approach and ignore these
contributions,\footnote{
By ignoring loops  one  assumes that the EFT is infinitely perturbative in the sense that any loop is  smaller than even the most irrelevant EFT operator. On the other hand, in this approximation, one
still keeps all orders in the EFT expansion.
This approximation is correct only in  the exact $g_{n,q} \to 0$ limit.
} were it not for the fact that \eq{eq:CHimp} is evaluated at $t=0$ where the loop effects entering into higher arcs, such as \eq{eq:blabla}, diverge.

Therefore, a finite~$t$ improvement procedure in the form of \eq{eq:geralimp}  rather than \eq{eq:CHimp} is qualitatively necessary to ensure the bounds survive at finite coupling.

\subsection{Improvement  at finite $t$} \label{subsec:FTIMP}
For the reasons outlined in the previous subsection, in what follows we introduce a procedure to improve arcs while not exploiting the information at $t=0$. 
This will eventually enable a perturbatively consistent derivation of bounds in theories with any IR structure $\Mh^{\text{loop}}\neq 0$.
Indeed, if we improve the \stl~arc of \eq{eq:CHimp} using only relations at $t\neq0$, the $\Mh^{\text{loop}}$ entering the improved arc will not be infinite when evaluated at $t\neq 0$, but  perturbatively small. It will then provide a genuinely small correction to the bounds, rather than compromising them.

To this goal, it is useful to express  derivatives of arcs in the form,
\begin{equation}\label{eq:darc} 
    \frac{1}{k!}\partial_t^k \, a_n(t) = \sum_{p=1}^{\infty} \, \sum_{q=0}^{p} \, \binom{p-q}{n+1-q} \binom{2 \, (p-n-1)+q}{k} \, t^{2\, (p-n-1)+q-k} \, g_{2p+q,q}\,,
\end{equation}
as derived in  Appendix~\ref{app:derIRarcs}.
We insert this expression into the definition of the improved arc given in \eq{eq:geralimp}, and require it to match the improved $a_0^{\text{imp.}}$ of \eq{eq:arcimpIR}. 
The result provides relations between the coefficients $c_{n,k}$.
For instance,  
\begin{align}\label{eq:partimpIR}
    a^{\text{tree}}_0(s,t)\!-\!2t^2a^{\text{tree}}_1(s,t)\!+\!t^3 \partial_t a^{\text{tree}}_1(s,t)\!-\!3 t^4a^{\text{tree}}_2(s,t)
      \!=\!-\frac{\kappa^2}{t}+g_{2,0}-g_{3,1}t +4 g_{8,0}t^6+\cdots
\end{align} 
eliminates the coefficients $g_{4,0},g_{5,1},g_{6,0},g_{7,1}$ from the 0th arc.
For each further $g_{2p+q,q}$ coefficient that we wish to cancel from $a_0^{\text{imp.}}$, there exists an additional constraint. The collection of these constraints leads to the following system of equations for the $c_{n,k}$'s:
\begin{equation}\label{eq:megasystem} 
    \sum_{n\geq0}\sum_{k\geq0} \, \binom{p-q}{n+1-q}\binom{2(p-n-1)+q}{k} \, c_{n,k}=0,\quad \forall g_{2p+q,q}\notin \mathcal{S}
\end{equation}
where $\mathcal{S}$ is the set of indices we wish to keep in the improved arc.
For instance, in order to improve the arc to the form of \eq{eq:arcimpIR}, the relevant set $\mathcal{S}$ is $\mathcal{S}=\left\{g_{2,0}, g_{3,1}\right\}$, which is equivalent to setting $c_{0,0}=1$ and $c_{0,1}=0$ in the system of equations. 
On the other hand, for \eq{eq:arcimpIR2} $\mathcal{S}=\left\{g_{2,0}, g_{3,1}, g_{4,0}\right\}$, such that  $c_{0,0}=1$, $c_{0,1}=0$ and $c_{1,0}=0$. 
In both cases  $c_{0,0}=1$ provides the correct normalisation because we start with the zeroth arc. 
In particular, the first $n$ arcs contain $(n^2+3n)/2$ couplings not appearing in higher arcs. 
Generically, solutions for some improved arc exist if there are at least $k\leq n+1$  derivatives per arc included in \eq{eq:geralimp}. This expectation does not hold in special cases, such as the improvement of \eq{eq:partimpIR}, where we require only $k\leq n$ derivatives per arc, due to accidental cancellations.
We can think of the solution with $k\leq n$ as the \emph{minimal one}, while solutions for higher $k$ are redundant, since they can be repackaged in the form of null constraints, as we show in Appendix~\ref{app:NCHA}.

The resulting algorithm is infinite, meaning that in order to cancel infinitely many coefficients, we need a series containing infinitely many arcs.
In practice, we will solve the system algorithmically after  \emph{truncating} \eq{eq:megasystem} at some value $n\leq
N$, which is equivalent to setting $c_{n,k}=0$ for $n\geq N$. 
Then \eq{eq:megasystem} becomes a linear system of equations with the unique solution,
\begin{eqnarray}\label{eq:explcnk}
    &&c_{1,0}= -2,\quad c_{1,1}= 1,\quad c_{1,2}= 0,\nn\\
    &&c_{2,0}= -3,\quad c_{2,1}= 0,\quad c_{2,2}= 1,\quad c_{2,3}= 0,\nn\\
    &&c_{3,0}= -8,\quad c_{3,1}= -2, \quad c_{3,2}= 2,\quad c_{3,3}= 1, \quad
    c_{3,4}= 0\nn\\
    &&\cdots
\end{eqnarray} 
which satisfies the recursion relation,
\begin{equation}\label{eq:recursion}
    c_{n+1,\, k}=c_{n,\, k-1}+2\, c_{n,\, k}+c_{n,\, k+1} \,.
\end{equation}
Alternatively, this solution can be expressed in a more compact form using the generating function,
\begin{equation} \label{eq:impossible} 
    G(x,y)\equiv
    \frac{x\left(1+\sqrt{1-4x}-6x\right)}{\sqrt{1-4x}\left[2x+y\left(2x-1+\sqrt{1-4x}\right)\right]}
    \,,
\end{equation}
as 
\begin{equation}\label{eq:coeffss}
    c_{n,k}= \frac{\partial_x^n \, \partial_y^k}{n!k!} \, G(x,y)\Bigg|_{x \, = \, y \, = \, 0} \,.
\end{equation}
In Appendix~\ref{app:MF} we give more details and a derivation of the partially improved arc including~$g_{4,0}$, as well as a more general algorithm to find other relations, such as the $t\neq 0$ analogue of null constraints.

Importantly, the improvement presented here is only  partial: if we include arcs only up to $N$, we cannot cancel coefficients such as $g_{2N+2,0}$, which require the arc~$a_N(s,t)$.
Following our discussion on  M\"untz-Sz\'asz, the use of this expression only makes sense if we postulate that the coefficients $g_{n,p}$ with $n>N$ are  bounded;
this way they can not reproduce the effects of the lower coefficients as implied by the M\"untz-Sz\'asz theorem, see Appendix~\ref{app:MS}.
This assumption is further justified by two facts that will be discussed in the next section. First
we will see that $|t|$ is limited by an upper value $t_\text{max}\ll s$, which implies that the neglected terms are exponentially small in $N$; secondly, 
the bounds derived with our algorithm  converge very quickly with $N$, and  asymptotically approach the ideal situation with no assumptions on the
coefficients.

\subsection{Partial improvement in the UV}\label{subsec:partimpuv}
Now that we have ensured that the problem is well posed, we can apply the same improvement algorithm to the UV representation of the arcs.  
This means substituting the UV arcs \eq{eq:archdeftUV2} in
the sum \eq{eq:geralimp}, obtaining,
\begin{equation}\label{eq:UVimpgenC} 
    a_0^\text{imp}(s,t) = \left\langle     I_{\ell} \left(\os,t \right) \right\rangle \,,
\end{equation} 
with
\begin{equation}
   I_{\ell} \left(\os,t \right) = \sum_{n,k} \, c_{n,k} \, \frac{t^{2 \, n+k}}{k!} \, \partial_t^k \, \frac{1}{[\os(\os+t)]^n} \,\frac{\left(2 \, \os+t\right)}{[\os(\os+t)]} \mathcal{P}_\ell \left(1+\frac{2\,t}{\os}\right) \, ,
\end{equation}
where the coefficients $c_{n,k}$ are given by \eq{eq:coeffss}. 
To our knowledge this function cannot be re-summed, unlike the improved arc in \eq{eq:CHimp} taken from Ref.~\cite{Caron-Huot:2021rmr}.  
Using a simple trick, we may nevertheless write it in such a way that some of its convergence properties are revealed to us.
We introduce an additional variable $\tilde{t}$ and bring the $t^{2n}$ factors in \eq{eq:UVimpgenC} inside the partial derivative by the replacement,
\begin{equation}
    t^{2\, n} \partial_t^k \left( \, \cdots \right)  =\partial_t^k \,\tilde{t}^{2\, n} \left( \, \cdots \right) \, \bigg |_{\tilde t = t} \, .
\end{equation} 
The sum in $n$ can then be carried out explicitly. 
The result is the $k$th order in $y$ of the  generating function \eq{eq:impossible}, with $x=\tilde t^2/s(s+t)$:
\begin{equation}\label{eq:UVimpgenC2} 
\begin{split}
    a_0^\text{imp}(s,t) &= \Big\langle \sum_{k} \frac{t^{k}}{k!}\partial_t^k \, \left(\frac{2}{\sqrt{1-\frac{4 \, \tilde t^2}{s (s+t)}}+1}-1\right)^k \\
    & \times \frac{ (2 \, s+t) \left(s (s+t) \left(\sqrt{1-\frac{4 \, \tilde t^2}{s (s+t)}}+1\right)-6 \, \tilde t^2\right) }{2(s (s+t))^{3/2} \sqrt{s (s+t)- 4 \, \tilde t^2}} \, \Bigg|_{\tilde t=t}\mathcal{P}_\ell\left(1+\frac{2t}{\os}\right) \Big\rangle \,.  
\end{split} 
\end{equation}
By inspecting the improved arc in this form, it is clear that the formula can not converge if the expression appearing under all of the square roots is negative, which puts a bound on the values of $t$ for which it can be used,
\begin{equation}\label{eq:tstar}
    -t_*\leq t \leq 0 \quad \textrm{with}\quad
    \frac{t_*}{s}\equiv\frac{1}{8} \left(\sqrt{17}-1\right)\approx 0.39\,.
\end{equation}
This is an important result, since it limits the range, $t_{\text{max}}\leq t_*$, over which the dispersion relations can be smeared against the measure appearing in \eq{eq:projection}.
The existence of the upper bound is supported by numerical results.

\subsection{Projection on basis functions}\label{subsec:projectionbasis}
Integrating both the IR improved arcs \eq{eq:arcimpIR} and their corresponding UV representations \eq{eq:UVimpgenC} against a complete set of basis functions $f_i(t)$ (as in \eq{eq:projection}), we obtain the most general set of IR-UV relations in algebraic form.
In the absence of gravity ($\kappa\to 0$), arcs are continuous functions of $t$.
For continuous functions, the space of polynomials provide a complete basis (Weierstrass theorem).  
Moreover, orthogonal polynomials---in particular Legendre polynomials $P_j$---are  useful as they provide a countable set for this goal.
Indeed, they are  orthogonal with respect to the flat $\dd t$ integration measure and therefore are the most natural choice of basis.

We consider a smearing function of the form,
\begin{equation}\label{eq:measure} 
    f(t)= \sum_{j=0}^{j_{\text{max}}} b_j P_j\Big(1+\frac{2t}{t_{\text{max}}}\Big)\,
\end{equation}
where in principle $j_{\text{max}}\to \infty$, but in practice it will be taken to be finite. 
Notice that when integrated against the IR representation of the arcs at tree-level, 
\begin{equation}\label{eq:onlyafew}
    \kappa\to0: \;\;\; \int_{-t_{\text{max}}}^0 \dd t \, f(t) \, a_0^{\text{imp}}(\bs,t) = b_0 \left(g_{2,0} \, t_{\text{max}}+g_{3,1}\frac{ t_{\text{max}}^2}{2} \right) -b_1 \,  g_{3,1} \, \frac{t_{\text{max}}^2}{6} \,, 
\end{equation} 
only the first two polynomials have support:
as the constant and linear terms in $t$ can be expressed entirely in terms of the first two Legendre polynomials, all $P_j$ with $j\geq 2$ integrate to zero on the IR side. 
Then, in a sense, the $j\geq 2$ polynomials provide the null constraints: when integrated against the UV representation of the arcs, the corresponding expression must vanish. 
For the improvement of \eq{eq:arcimpIR2}, which includes $g_{4,0}$, the same holds true for~$j\geq 3$.

In the presence of gravity, the IR arcs are no longer analytic in $t\in [\,-t_{\text{max}},0 \, ]$.
Nevertheless, by modifying the integration measure to
\begin{equation}\label{eq:measuregrav}
    \dd\mu(t)\equiv (-t/t_{\text{max}})^\alpha f(t)\,,\quad \textrm{with}\quad
    \alpha>0\,, 
\end{equation}
the arcs become integrable. 
Continuous functions span the set of integrable functions on an interval. 
Therefore polynomials, which are themselves a basis of continuous functions, again provide a complete basis. 
In this work we take $\alpha=1$, in order to preserve the property that, at tree-level, higher Legendre polynomials integrate to $0$ in the IR and provide null constraints. 
On the other hand, the small $j$ terms have support in the IR leading to,
\begin{equation}\label{eq:onlyafew2}
\begin{split}
    \kappa\neq 0: \;\;\; \int_{-t_{\text{max}}}^0 \dd t\,\,\left(- t/t_{\text{max}} \right)\,& f(t) \,  a_0^{\text{imp}}(\bs,t) =\,\, b_0 \, \left(\kappa^2 +g_{2,0} \, \frac{ t_{\text{max}}}{2}+g_{3,1} \, \frac{ t_{\text{max}}^2}{3} \right) \\
    &-b_1 \, t_{\text{max}} \, \left(\frac{g_{2,0}}{6} +g_{3,1} \, \frac{t_{\text{max}}}{6}\right) +b_2 \, g_{3,1} \, \frac{t_{\text{max}}^2}{30} \,. 
\end{split} 
\end{equation}

Notice that in this expression the integral with $j=2$ is non-vanishing, as opposed to the case without gravity of \eq{eq:onlyafew}.
\vspace{5mm}

Now we need to compute the integrals on the UV side, as in \eq{eq:UVimpgenC}. 
We express the smeared \eq{eq:UVimpgenC} as
\begin{equation}\label{eq:UVintsdone}
      \int_{-t_{\text{max}}}^0 \dd t\, \left(-t /t_{\rm max}\right)^\alpha \, f \left(t \right) \, a_0^{\text{imp}}(\bs,t) = \int_{-t_{\text{max}}}^0 \dd t\, \left(-t /t_{\rm max}\right)^\alpha \, f \left(t\right) \Big\langle I_{\ell} \left(s^\prime, t \right) \Big\rangle\,,
\end{equation}
where we can take $\alpha=0$ in the absence of gravity, and $\alpha=1$ otherwise.
We then assume convergence and exchange the various orders of integration and sums.
It turns out that for all fixed values of $k$ and $n$ for the coefficient $c_{n,k}$, and for each value of $\ell$, the integrals in $t$ can be performed analytically.

In summary, the equalities between UV expressions (\ref{eq:UVintsdone}), and smeared IR arcs (which in the \stl~approximation look like \eq{eq:onlyafew}, or \eq{eq:onlyafew2} with gravity) allow one to extract bounds on Wilson coefficients. 
This means that we can access the information contained in dispersion relations at finite~$t$, through the space of basis functions spanned by the coefficients $b_j$ in \eq{eq:measure}.
We will show this explicitly in the next section.

\section{Bounds on Wilson coefficients}\label{sec:boundswc}

The finite~$t$ dispersion relations obtained in Section~\ref{sec:finitetimprovement} imply bounds on the IR Wilson coefficients deriving from the positivity properties of the average  $\langle \, \cdots \rangle$ of \eq{eq:average}. 
In the forward limit the bounds can be obtained analytically and at all orders by expressing the UV integrals as moments~\cite{Bellazzini:2020cot}.  
As for the smeared dispersion relations discussed above, there is no similar formulation, but the problem can be recast as a linear or semi-definite optimisation problem and solved numerically, as proposed in Ref.~\cite{Caron-Huot:2021rmr}.
We take this approach in what follows.

We have the freedom to choose the parameters $b_j$  of the smearing function $f(t)$ to ensure that the projected UV representation is positive. 
Assuming convergence, we take the $t$ integral inside the other sums and integrals, where it can be computed explicitly.
We then demand positivity of the integrand for all values of $\ell,\os$:
\begin{equation}\label{eq:statement_semidefinite}
     \textrm{UV:} \;\; \int_{-t_{\text{max}}}^0 \dd \mu(t) \, I_{\ell}^{N}\left(s',t\right) \geq 0, \;\;  \forall \, \os \, \textrm{and} \;\ell \; \Rightarrow  \; \textrm{IR:} \;\; \int_{-t_{\text{max}}}^0  \dd \mu(t)\, a_{0}^{\text{imp}, N}\left(t\right) \geq 0, 
\end{equation} 
where we use the notation $I_\ell^{N}$ and $a_{0}^{\text{imp}, N}\left(t\right)$  to refer to the truncated sum and arc used for numerical applications. 
Here  $a_{0}^{\text{imp}, N}\left(t\right)$ is the truncated, improved arc \eq{eq:geralimp} in the IR representation (as in \eq{eq:arcimpIR} or \eq{eq:arcimpIR2}). 

Therefore, if we can ensure positivity of the $t$ integrals of each individual $I_{\ell}^{N}(s',t)$, it implies a positivity statement for the IR improved arc.
In practice, we truncate various parameters: the maximum order in $t_\text{max}$ by which the arc is improved, the maximum order in the partial wave expansion, and the maximum order of the smearing polynomials.
This introduces three parameters: $N$, $\ell_\text{max}$, and $j_\text{max}$, respectively.

On the other hand, to ensure positivity in $\os$, differently from Ref.~\cite{Caron-Huot:2021rmr}, we approximate the integrand in $\os$ with a polynomial, for every $\ell$. 
Such an approximation follows naturally from  expanding in $t_{\text{max}}/\os$ as a Taylor series, because $\os/s>1$ from the integral range and $t_{\text{max}}/s\ll 1$ from kinematics. 
For consistency with truncation, we keep terms up to order~$2N$ in~$t_{\rm max}$.
This approach is advantageous because positive polynomials are simpler to characterise than positive functions: every positive polynomial can be written as a sum of squares of polynomials, which have a straightforward characterisation in terms of semi-definite positive matrices.
Furthermore, for every fixed $\ell$, it captures all values of $\os$.

In Appendix~\ref{app:smerding} we outline the technical details of how this is implemented in practice, quickly reviewing Ref.~\cite{Caron-Huot:2021rmr} and highlighting the differences w.r.t.\ our approach.

\subsection{Bounds at $\kappa=0$}\label{subsec:NoGravity}
In Section~\ref{subsec:partimpuv}, we pointed out that our partial improvement algorithm has a finite radius of convergence, which requires smearing over $-t_*<t\leq 0$, see \eq{eq:tstar}.
In order to compare more easily with previous literature, in this section we present results using both our approach and the one of Ref.~\cite{Caron-Huot:2021rmr}, for different  smearing windows $-t_{\text{max}}\leq t\leq0$.
\begin{figure}[h]
    \centering
    \includegraphics[width=0.7\textwidth]{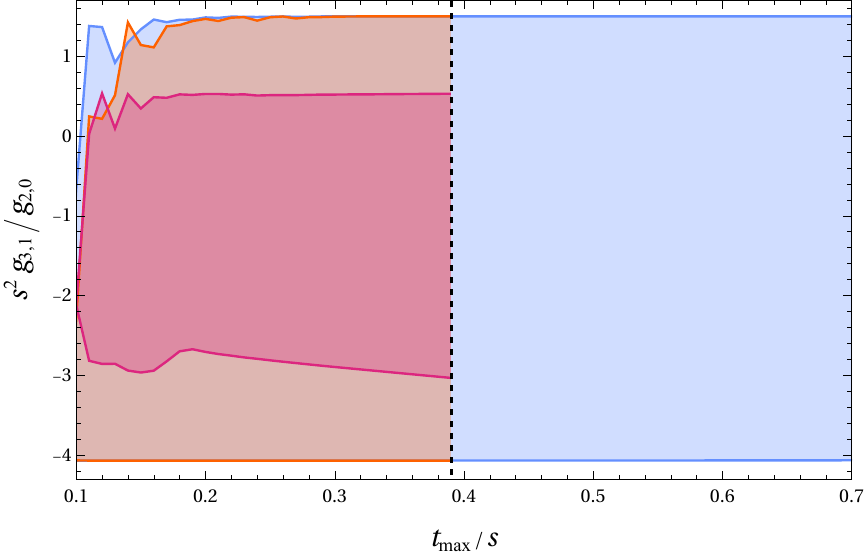}
    \caption{\it\label{fig:tmaxdependence}
    Bounds on the ratio $g_{3,1}s/g_{2,0}$ as a function of $t_{\text{max}}$, the maximum value of $t$ in the smearing integral. All curves are derived with fixed $j_\text{max}=7$, $\ell_\text{max}=14$.
 In blue the improvement approach of Ref.~\cite{Caron-Huot:2021rmr} where positivity is obtained expanding the integrand at order $O(t_\text{max}^{16})$. In purple (orange), our improvement at order $N=3$ ($N=8$). The dashed vertical line delimits the applicability of our improvement formula,~$t_\text{max}\leq t_*$. As $N$ is increased, the algorithm at finite~$t$ of Section~\ref{sec:finitetimprovement} rapidly converges to the value of~Ref.~\cite{Caron-Huot:2021rmr}.}
\end{figure}

We first work in the $\kappa \to 0$ limit and $d=6$  for easier comparison with the next section. Bounds in $d=4$ share the same qualitative features, although the numerical values for the lower bound vary slightly.

In Fig.~\ref{fig:tmaxdependence} we show upper and lower bounds on the ratio $g_{3,1}/g_{2,0}$ in units of the arc energy scale $s$, for different values of $t_\text{max}$, fixing the parameters  $j_\text{max}=7$, $\ell_\text{max}=14$ used in the bound derivation. The blue line is obtained using the algorithm of Ref.~\cite{Caron-Huot:2021rmr}, but expanded in the UV side in powers of $t_\text{max}/s$ (hence the plot ranging only to $t_\text{max}=0.7s$). In purple/orange, instead, our approach for $N=3,8$ respectively, limited to $t_\text{max}<t_*$ (dashed black line). As discussed previously, these correspond to neglecting terms of order $O(t_\text{max}^{7})$ and $O(t_\text{max}^{17})$.

\begin{figure}[h]
    \centering
    \includegraphics[width=0.7\textwidth]{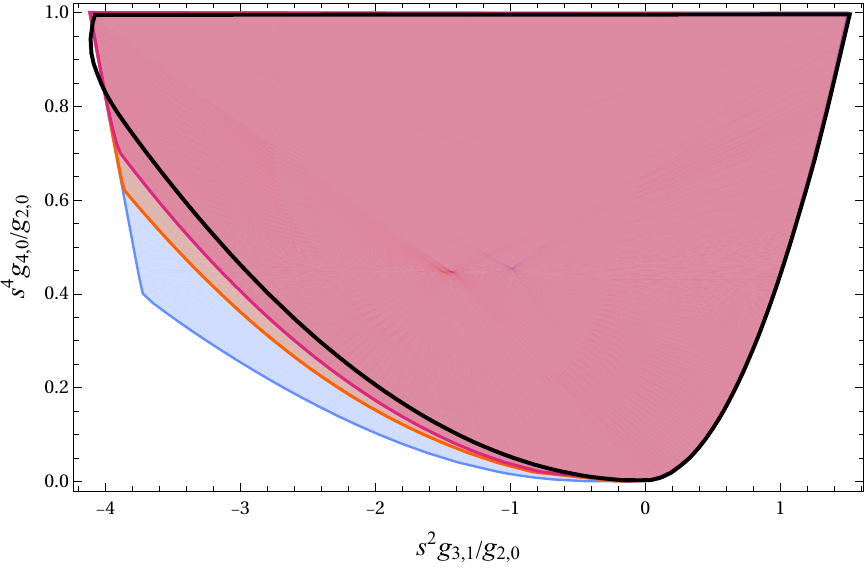}
    \caption{
    \it Positivity bounds on the ratio of coefficients $g_{4,0}/g_{2,0}$ and $g_{3,1}/g_{2,0}$ in units of $s$, for $d=6$, using $\ell_\text{max}=100$ and $j_\text{max}=10$.
    In blue the bounds using our partial improvement \eq{eq:UVimpgenC} with $N=6$ and $t_{\text{max}}=t_*/2\approx 0.2 s$. 
    The approach of Ref.~\cite{Caron-Huot:2021rmr} is shown for comparison in black for $t_{\text{max}}=s$, in red for $t_{\text{max}}=0.8s$ and orange for $t_{\text{max}}=0.5s$.\label{fig:g3g4}}
\end{figure}

From Fig.~\ref{fig:tmaxdependence}   we conclude that our algorithm correctly reproduces the 1-dimensional bounds without gravity, despite the limit on the smearing range $t_\text{max}<t_*$. Furthermore, we notice that it rapidly converges with $N$. In particular, for $N=8$ the obtained bounds are indistinguishable from the exact results. Finally
for fixed $j_\text{max},\ell_\text{max}$, the bounds become unreliable as $t_\text{max}\to 0$.

In Figure~\ref{fig:g3g4} we present a similar analysis, extended to bounds on the ratio of coefficients $g_{3,1}s/g_{2,0}$ and $g_{4,0}s^2/g_{2,0}$, using $\ell_\text{max}=100$, $j_\text{max}=10$. 
First, we explore how changing the lower bound of integration \eq{eq:projection} affects the bounds in  the approach of Ref.~\cite{Caron-Huot:2021rmr}. The black line, the orange and the red areas are smeared  over  $-t_{\text{max}}<t\leq 0$ with $t_{\text{max}}=s$ in black, $t_{\text{max}}=0.5s$ in orange, and $t_\text{max}=0.8s$ in purple.
For $t_\text{max}= s$ the figure coincides with the forward limit bounds~\cite{Caron-Huot:2020cmc,Chiang:2021ziz}.
We observe that constraints corresponding to the algorithm of Ref.~\cite{Caron-Huot:2021rmr} become weaker as we take smaller values of $t_\text{max}$.

The blue area shows instead our approach, with the smearing range $t_\text{max}=0.2s$.  
Our partial improvement goes up to order $N=6$, meaning we neglect terms of order $O(t_\text{max}^{12})$ in the improvement, which, given the value of $t_*$, are of order $10^{-8}\times g_{2n,q}s^{2n}$.
We checked that the results are stable for larger values of $N$, meaning that the truncated algorithm has converged.

In Figure~\ref{fig:g3g4}, for both algorithms, we observe a feature that develops as soon as $t_\text{max}< s$: a kink in the lower bound. 
The fact that for large values of $g_{4,0}s^2/g_{2,0}$ the $t_\text{max}<s$ bounds are slightly more stringent than the $t_\text{max}=s$ ones, makes us think that this feature is a numerical artifact.
Were this not the case, one would expect that even for $t_\text{max}=s$ the test polynomials would arrange to (almost) vanish at large $|t|$, to obtain a more stringent result. 
On the other hand, we have tried several approaches to improve the bounds in this region.
We have explored values as large as $\ell_\text{max}=600$ and introduced the $\ell\to\infty$ constraints discussed in Ref.~\cite{Caron-Huot:2021rmr}.
Moreover, we have produced  the red curve by employing two different methods. 
In the first we have followed exactly the procedure detailed in Ref.~\cite{Caron-Huot:2021rmr}, where positivity of \eq{eq:statement_semidefinite} is ensured by discretising $\os$ and requiring that the integrand be positive at each discrete value.
In the second we have used the expansion of the integrand in $t_\text{max}$, and its polynomial approximation in $\os$, as described above. 
Despite having explored various approaches and increased numerical precision, the kink on the left hand side of Figure~\ref{fig:g3g4} survives.
Regardless of this feature, the results of our algorithm are qualitatively consistent with those obtained from the approach of~\cite{Caron-Huot:2021rmr}.

\subsection{Bounds at $\kappa \neq 0$}
In this section we work in the presence of gravity, at $\kappa\neq0$, and in $d=5,6$ dimensions, where the partial wave expansion converges and the $ 2\to 2 $ amplitude is well defined w.r.t.\ soft graviton emission.

\begin{figure}[htb]
    \centering
    \includegraphics[width=0.7\textwidth]{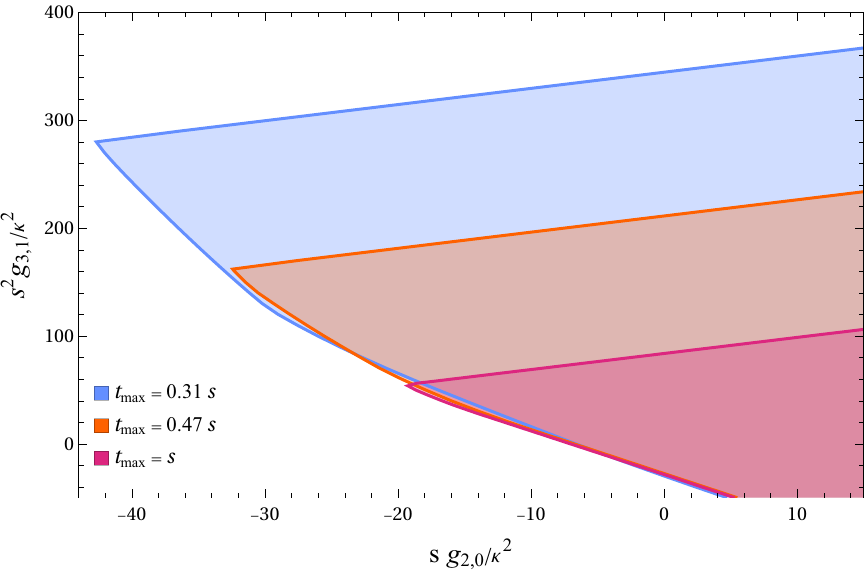}
    \caption{\it \footnotesize Bounds in the presence of gravity on the ratios $g_{2,0}s/\kappa^2$ and $g_{3,1}s^2/\kappa^2$, in $d=6$, for different values of $t_{\rm max}$. All bounds in this figure are derived using the improvement of  Ref.~\cite{Caron-Huot:2021rmr}.
    The blue (orange) line uses $j_\text{max}=6\,\,(9)$, and is expanded up to order $16\,\,(25)$ in $t_\text{max}$.
    }
    \label{fig:gravtmax}
\end{figure}

\begin{figure}[h]
    \centering
        \includegraphics[width=0.7\textwidth]{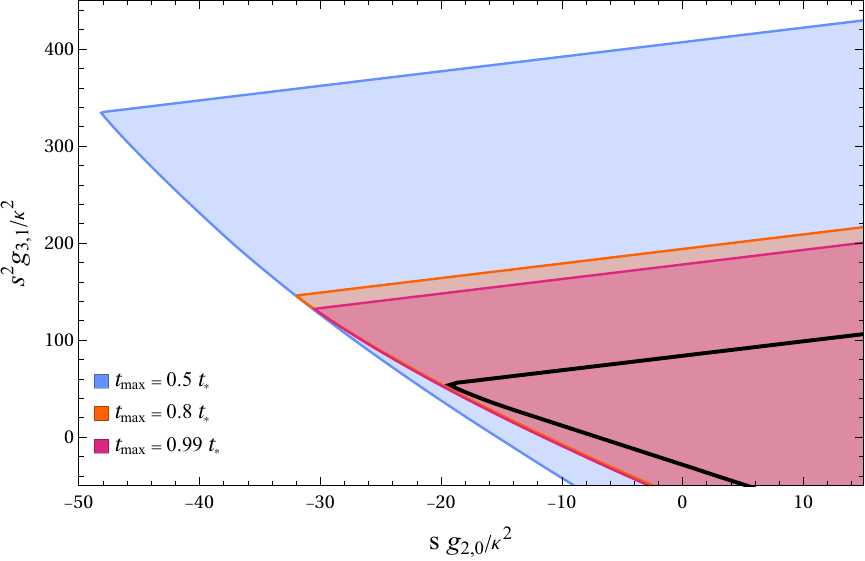}
    \caption{
    \it \footnotesize Same as Figure~\ref{fig:gravtmax}, but  using our improvement formula \eq{eq:UVimpgenC} up to order $N=12$ (refer to Table~\ref{tab:consistentparams} and~\ref{tb:values} for the exact choices of parameters we make).  All the bounds are computed with $\ell_\text{max}=400$, and up to $j_\text{max}$=8.  Here $t_{\max}$ is expressed in units of the maximum allowed smearing range $t_*\approx  0.39 s $. For reference, in black we show the bounds from  Ref.~\cite{Caron-Huot:2021rmr}. 
    }\label{fig:gravd6}
\end{figure}

In Figures~\ref{fig:gravtmax} and~\ref{fig:gravd6} we show the bounds on the coefficients $g_{2,0}$
and $g_{3,1}$ in units of the gravitational coupling $\kappa^2$ and the arc energy scale $s$. Both figures are derived in $d=6$ dimensions and use $\ell_{\rm max}=400$, and up to $j_\text{max}=9$.
Figure~\ref{fig:gravtmax} uses  the method of Ref.~\cite{Caron-Huot:2021rmr}, but for different values of $t_{\text{max}}$. Figure~\ref{fig:gravd6} uses our method \eq{eq:UVintsdone}, 
where the smearing polynomials involve the overall $(-t)^\alpha$ factor, with $\alpha=1$, in order to make the EFT amplitudes integrable (see Appendix~\ref{app:smerding} for more details).

\begin{figure}[h]
    \centering
    \includegraphics{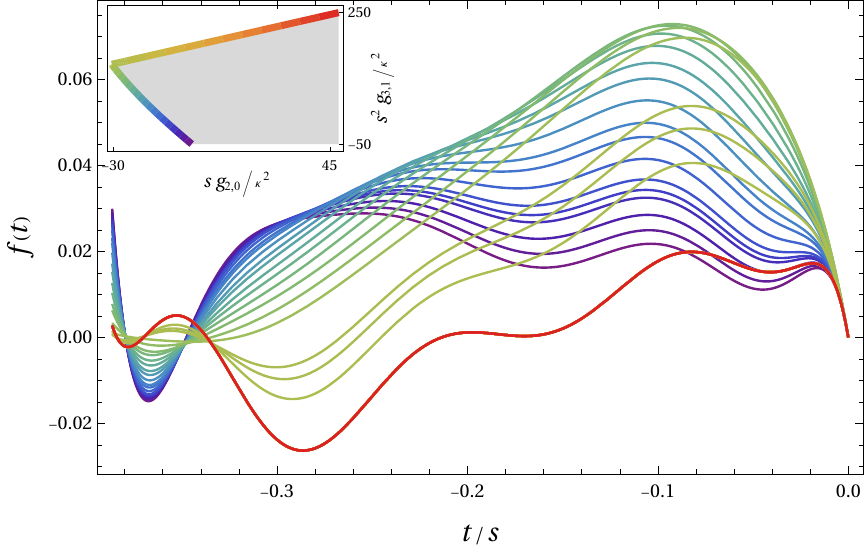}
    \caption{In the top left-hand corner, the $d=6$ bound of Figure~\ref{fig:gravd562} for $t_{\rm max}=0.99t_*$, with each extremal point color-coded.
   In the main figure,  for each extremal point, the corresponding  smearing functions $f(t)$ as a function of $t$.
    The functions are normalised as $\sum _{j=0}^{j_{\rm max}}b_j=1$. Here, we use $j_\text{max}=8$.
    }\label{fig:functionals}
\end{figure}

In Figure~\ref{fig:functionals}, we show the $t$ dependence of the functionals $f(t)$ that optimize our bounds, for different extremal points, defined by the colours in the top left-hand corner plot.
Gravity plays the most important role for small values of the couplings.
Interestingly, there, the functionals exhibit a distinctive peak at finite values of $t$, far from the maximum allowed value $t_{\rm max}$, while staying away from $t=0$ to avoid the gravity pole.
The existence of a scale associated with this feature provides an explanation of why the  bounds become weaker as the smearing range to $t_{\text{max}}<s$ is reduced. For larger values of the couplings, the functionals morph into oscillatory functions, which are less impacted by the $t_{\rm max}$ cut. Indeed the asymptotic  upper and lower slopes $g_{3,1}/g_{2,0}$ are almost insensitive to $t_{\rm max}$, as already observed in Figure~\ref{fig:tmaxdependence}.
 
\begin{figure}[h]
    \centering
    \includegraphics[width=0.7\textwidth]{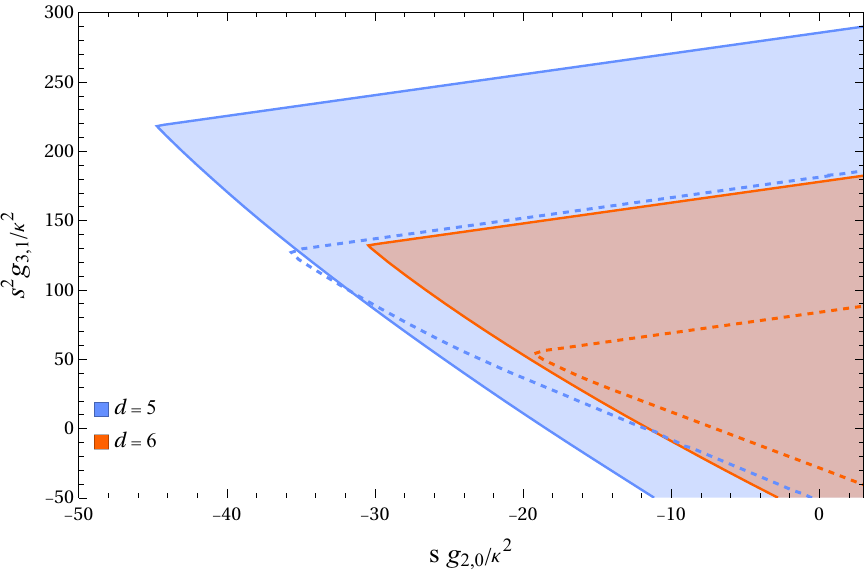}
    \caption{
    \it \footnotesize 
    The same bounds as in Figure~\ref{fig:gravd6}, in $d=5$ (blue) and $d=6$ (red) dimensions. The solid lines are derived using our method, while the dashed ones are taken from Ref.~\cite{Caron-Huot:2021rmr}.\label{fig:gravd562}}
\end{figure}

Finally, in Fig.~\ref{fig:gravd562} we apply our method to find bounds in the presence of  gravity in $d=5$ (blue) and $d=6$ (red) dimensions. For comparison, the dashed lines are taken from Ref.~\cite{Caron-Huot:2021rmr}.
Although what first hits the eye is the fact that our bounds  are less stringent, they are in a sense more robust as they will be stable  
against the introduction of IR loop effects.
Moreover, as discussed above, they are penalised by the $t_{\rm max}<s$ constraint. It is plausible that a variation of our algorithm exists  that  possesses a larger radius of convergence, and which re-sums to all orders. In the outlook and Appendix~\ref{sec:apppol} we discuss a promising direction to find such an algorithm, the thorough study of which we leave for future work.

\FloatBarrier

\section{Conclusions and outlook}\label{sec:conclusion}

Positivity bounds shape the  swampland into the subset of theories consistent with UV completions, whether weakly or strongly coupled. Despite the fact that most results in this field have been derived via Taylor expanding UV-IR dispersion relations in the vicinity of the forward limit, this approach fails in IR theories with gravity  or quantum effects associated with loops of massless particles~\cite{Bellazzini:2021oaj,gravityloops}.

In this work we have presented a procedure that works by entirely avoiding the forward limit, adding an important ingredient to Ref.~\cite{Caron-Huot:2021rmr}. It provides a proof of principle that positivity bounds apply in all theories, independently of their IR structure. It is based on a combination of arcs (dispersion relations with different subtractions), at $t\neq 0$. In the UV, this can be integrated in $t$ against some weight function, to a manifestly positive quantity. In the IR this combination of arcs will have non analyticities from gravity poles and loop discontinuities, but it will be integrable into a finite quantity. 

In other words,  this work opens the door to treat loop effects as perturbatively small, rather than as an obstruction to positivity bounds.

Contrary to results in the strictly tree-level approximation, which can be packaged into (sometimes exact and analytic) relations among any finite numbers of coefficients or arcs,  the results found here are based on approximating all infinite integrals and sums into finite ones. In this sense, our approach is rooted in the spirit of perturbation theory: it works at finite (albeit arbitrarily large) order in the EFT expansion and can work at any order in the perturbative loop expansion.
Loop effects introduce infinitely many terms scaling with higher orders in $t$ anyway, and these cannot be removed by improvement. 
In light of this, partial improvement is qualitatively as good
as it can be.

While discussed in the context of gravity, our approach is designed to work in any theory with massless particles. It would be interesting to study the IR structure in these theories in detail and evaluate their impact on positivity bounds quantitatively. In particular in gravity \cite{Caron-Huot:2022ugt},  spin-1 bosons~\cite{Henriksson:2021ymi, Henriksson:2022oeu, Bertucci:2024qzt} or  pion physics~\cite{Distler:2006if,Manohar:2008tc,Albert:2022oes,Fernandez:2022kzi}
to understand for instance the interplay between these effects and bounds on the chiral anomaly~\cite{Albert:2023jtd,Ma:2023vgc}. We leave this for future work~\cite{gravityloops}.

It would be interesting to see if some of our results can be re-summed into compact expressions valid at all orders.
Such expressions might manifestly reveal properties of the UV extreme amplitudes that saturate the bounds.
The \emph{partial improvement} procedure we proposed is only one of many: it is based on identifying the smallest set of arcs needed to cancel all terms, but other combinations are also possible. Alternatively, adding the null constraints, as described in Appendix~\ref{app:NCHA}, could lead to more compact forms for the improved arc. Manifestly crossing symmetric dispersion relations~\cite{Sinha:2020win}, might provide a different take on this problem.

An alternative formulation, which we describe in more detail in Appendix~\ref{sec:apppol}, is to combine arcs with polynomials~$p_n(t)$,
 \begin{equation}\label{eq:geralimpPOL} 
    a^\text{imp-pol}(s,t) =
    \sum_{n}p_n(t)a_n(s,t)\,, 
\end{equation} 
in such a way that  the
\emph{integrated} IR arc contains only the coefficients we are
interested in bounding, e.g.,
\begin{equation}\label{imrpvementpolynomial}
    \int_{-t_{\text{max}}}^0 \dd t a_0^\text{imp-pol}(s,t)=\alpha {\kappa}+ \beta
    g_{2,0}+\gamma g_{3,1}\,.
\end{equation} 
Here
$\alpha,\beta,\gamma$ are some parameters associated with the normalisation of
said polynomials, and that eventually play the role of floating coefficients in
the semi-definite optimization problem.
In practice, in the IR  improved arc, each
higher coefficient $g_{4,0},g_{5,1},\cdots$ multiplies a polynomial in $t$ that
integrates to zero, but which itself is not zero. 
The advantage of this method 
is that it does not involve derivatives of arcs, and is therefore more in line with the idea that dispersion relations are exact in the distributional sense.

In this article we have stressed the importance of the M\"untz-Sz\'asz theorem, which highlights the tension between   Taylor expansions and their 
appearance into distributions.
Besides our case of smeared positivity bounds, such a situation generally arises  in the context of~EFTs.
Indeed, the very idea of EFT is based on Taylor expanding  around small energies.
The amplitude stemming from such an EFT is then used as a distribution in  cross sections, where it is  integrated over a finite energy range, or a certain measure.
The M\"untz-Sz\'asz theorem implies that to reverse-engineer this, i.e.\ to extract the coefficients from integrated distributions, we need an infinite amount of extra information. This can be in the form of assumptions about the convergence of the EFT, as is the case for phenomenological applications of EFTs~\cite{Contino:2016jqw}, or in the form of null constraints as implied by full crossing symmetry at tree-level, as for positivity bounds.
This problem could also potentially appear in the context of the non-perturbative S-Matrix bootstrap~\cite{Paulos_2017,Paulos:2016but}, where one searches for a bound on the low-lying coefficients in the energy expansion of a generic function; it would be interesting to understand whether analyticity in both $s$ and $t$ provides a solution to this puzzle.

\section*{Acknowledgments}
We thank B.~Bellazzini, K.~Haring, D.~Karateev, D.~Kosmopoulos, M.~Marino, J.~Parra-Martinez, M.~Riembau, M.~Tacchi, A.~Zhiboedov for interesting discussions.
The work of C.B. was supported by Swiss National Science Foundation (SNSF) Ambizione grant PZ00P2-193322. G.I. is supported by the US Department of Energy under award number DE-SC0024224, the Sloan Foundation and the Mani L. Bhaumik Institute for Theoretical Physics.
G.I., D.P., S.R. and F.R. have been supported by the SNSF under grants no. 200021-205016 and PP00P2-206149.

\appendix

\section{M\"untz-Sz\'asz in practice}\label{app:MS}
The M\"untz-Sz\'asz theorem provides a fundamental obstruction to reconstruct the individual  coefficients in a Taylor expansion when all the information at our disposal is in the form of its integrals against test measures.

Now, while the fundamental obstruction is indeed there, in this section we want to make its implications more quantitative. Given a series expansion $\sum_n g_n \, t^n $, we   ask how would  the ``M\"untz-Sz\'asz alternative solution'' 
$\sum_{\lambda_n} \tilde g_n \, t^{\lambda_n} $
look like if the basis were missing one or more of the monomials, for instance $\lambda_n \notin [n_-,n_+]$ for some $n_+\geq n_- \geq 0$. More explicitly, we want to approximate the function $t$, with higher powers  from $t^{2 \, N}$ to some other $t^{N_{\rm max}}$, to some precision $\epsilon$,
\begin{equation}\label{approxMSeq}
    \left| t - \sum_{n=2N}^{N_{\rm max}} \tilde g_n \, t^{n}\right| < \epsilon\,,\quad t\in[0,1]\,.
\end{equation}
This is exactly the situation we are focusing on in the main text, where  the function $t$ is associated to the coefficient $g_{3,1}$ which  we are aiming at,
and we have removed all terms up to $t^{2 \, N}$ in the partially improved arc; $t^{N_{\rm max}}$ would be the highest term that can be resolved by the computer.

\begin{figure}[h]
    \centering
    \includegraphics[width=0.7\textwidth]{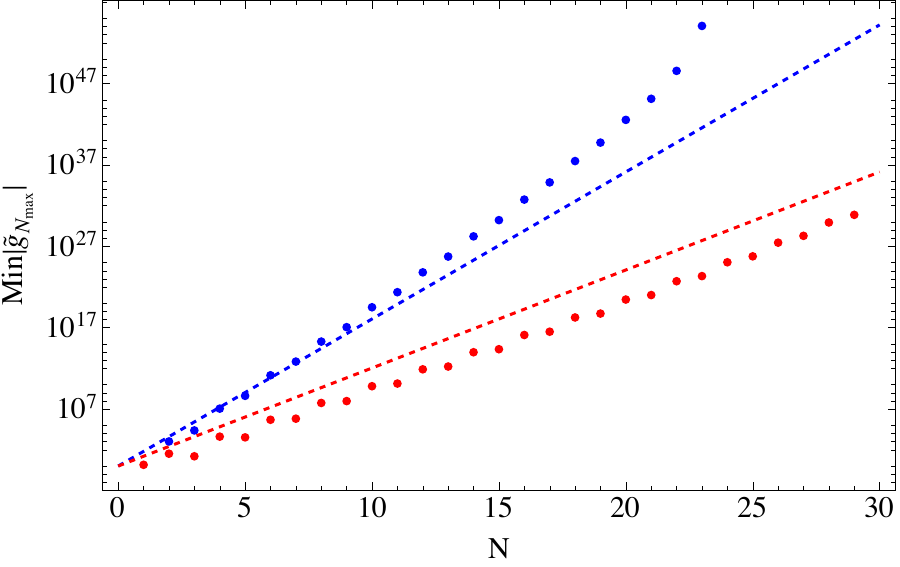}
    \caption{The size of the  smallest possible coefficient $g_{N_{\rm max}}$ in order to fulfil \eq{approxMSeq}, with $N_{\rm max}=4 \, N$ and $\epsilon=1/10$ (red) or  $\epsilon = 1/20$ and $N_{\rm max} = 6 \, N$ (blue). The  minimal coefficient was found solving a simple linear programming algorithm in \texttt{Mathematica}, by discretising $t$ over 100 points in the interval $[0,1]$ and imposing \eq{approxMSeq} in each point. The dashed red (blue) line corresponds to $g_{N_{\rm max}}\sim 2^{N_{\rm max}}$ for $N_{\rm max}=4 \, N(6\, N)$. }\label{fig:MS}
\end{figure}

In figure~\ref{fig:MS} we show the minimum (absolute) value for the last coefficient in the series~$|\tilde g_{N}|$, as a function of $N$, in two cases. In the first we demand $\epsilon = 1/10$ and allow $N_{\rm max} =4 \, N$ (red dots). In the second (blue dots) we take $\epsilon = 1/20$: in this case the previous value of~$N$ was not enough to approximate the function to the desired accuracy, and we had to take more terms in the series: $N_{\rm max} = 6 \, N$. Even then, for $N>22$ the problem \eq{approxMSeq} has no solution, and a higher $N$ should have been taken.

Ref.~\cite{Trefethen:2023:SLE} studied a similar problem, the approximation of $\abs{t}$ in $t\in[0,1]$ with a basis of even-power monomials $t^{2k}$, with $k=1,\cdots,n$. The author proves that for the approximation to be valid with precision $\epsilon$, the series requires at least $2n=1/(20 \epsilon)$ terms, and the coefficients are as big as $( 0.75 \, \epsilon ) \,2^{1/(40 \epsilon)}$, implying coefficients as large as $10^{107'000}$ for an $\epsilon=10^{-6}$ precision.

So, these results tell us two things about how the M\"untz-Sz\'asz theorem is realised in practice.
First of all, the coefficients can become so large, that it is plausible that in  numerical studies the ``M\"untz-Sz\'asz alternative solution'' discussed in the main text cannot be found. 
Secondly, and most importantly, it also tells us that even the weakest assumptions on the behaviour of the coefficients as $n$ grows, are enough to avoid  the solution taking over. 
For reference, the dashed  lines in Fig.~\ref{fig:MS} correspond to coefficients that grow as $g_{N_{\rm max}}\sim 2^{N_{\rm max}}$ in the two scenarios above (where $N_{\rm max}=4 N(6N)$) -- these would be the coefficients  in an  EFT with ordinary power-counting if  the ratios of energy versus cutoff were of order 2, rather than being much smaller than one.
Furthermore, from a physical point of view, it is reasonable to assume that after a certain order, the coefficients start decreasing, otherwise the whole idea of EFT loses its appeal. 
This assumption would be enough, by far, to prevent the higher coefficients from playing the role of the smaller ones and invalidating the bounds.

\section{Partial improvements at finite $t$}\label{app:MF}

\subsection{A compact form for IR arcs and their derivatives}\label{app:derIRarcs}

In the \stl~approximation and without gravity $\kappa\to 0$, the IR representation of arcs can be written in a compact form via \eq{eq:amptreeXY},
\begin{align} 
    a_n(t) &= \oint\frac{\dd
    s'}{2\PI\I} \frac{\mathcal{M}^{\text{tree}}_{\text{EFT}}(s',t)}{(s')^{n+2}(s'+t)^{n+1}}=
    \sum_{p\geq1}\sum_{q\geq0}g_{2p+q,q}(-t)^q \oint\frac{\dd s'}{2\PI\I}
    \frac{\left[(s')^2+t^2+s't\right]^{p-q}}{(s')^{n-q+2}(s'+t)^{n-q+1}}\nonumber\\
    &= \sum_{p\geq1}\sum_{q\geq0}g_{2p+q,q}(-t)^{2(p-n-1)+q} \alpha_{p-q,\,n-q+1}.
    \label{eqarcsallord}
\end{align} 
In the last step, we have pulled an overall $t$-dependent factor outside of the integral and have used,
\begin{equation}
    \alpha_{n,m}= \oint\frac{\dd z}{2\PI\I}
    \frac{(z^2-z+1)^n}{z^{m+1}(z-1)^m}=\binom{n}{m},
\end{equation}
where the integral is easily solved using the binomial and residue theorems.
Using this, we can then take $k$ derivatives in $t$ and obtain \eq{eq:darc}.

\subsection{Null constraints and higher-arcs}\label{app:NCHA}
Here we provide a general procedure for improvement and to find null constraints algorithmically, without using  arcs (or their derivatives) in the forward limit.  
It simply consists in gradually removing all the coefficients $g_{n,q}$ from an initial object, using higher arcs. The coefficients are dimensionful and they always multiply the same powers of $t$ when grouped into dimensionless objects. For simplicity we neglect gravity in this section. Its inclusion is trivial.

The starting point is to identify an object to be improved. In the main text Section~\ref{subsec:FTIMP} this is $a_0(s,t)$, but we might as well start from $a_1(s,t)$ to improve the first arc, or we could start from $\partial_t ^4 a_1(s,t)$: in this case we could use higher arcs to remove all coefficients, and we would obtain genuine null constraints.\footnote{Null constraints can also be obtained trivially by taking $m+1$ derivatives in $t$ in an arc $a_{n\geq1}(s,t)$ improved up to order $m$.} We will follow here the example of $a_0$ relevant for the main text, in a way that will make its generalisation straightforward. So, using \eq{eqarcsallord}, the object to improve is,
\begin{equation}\label{a0appe}
  a_0(s,t)=  \sum_{p\geq1}^{N}\sum_{q\geq0}^{N-n}\binom{p-q}{1-q}g_{2p+q,q}(-t)^{2(p-1)+q} \,,
\end{equation}
where we have highlighted that, although the original arc has infinitely many terms, the procedure is constructed to cancel a finite number of them, up to $N$.
We have the choice to truncate in both the number of arcs used and also at some order in $t$; in practice we choose these to be the same.

To improve this, we can subtract combinations of arcs $a_n$, with $n\geq1$, and their derivatives~$\partial_t^k a_n$, where $0\leq k\leq n+1$ is in principle necessary to find a solution:
\begin{equation}
  a_0^{\text{imp},N}=\sum_{n=1}^{N}\sum_{k=0}^{N-n} c_{n,k}
    \sum_{p\geq1}\sum_{q=0}^{n+1}
    \binom{p-q}{n+1-q}\binom{2(p-n-1)+q}{k}t^{2(p-1)+q}g_{2p+q,q}\,,
    \label{eq:darc2} 
\end{equation}
which is simply \eq{eq:geralimp} combined with \eq{eq:darc} for the arc's $k$-th derivatives.
The sum of these two expressions \eq{a0appe} and \eq{eq:darc2}, must give the improved arc $g_{2,0}-g_{3,1} t$. 
We can phrase this into matrix notation by defining a vector $\vec g\equiv \{g_{2,0},g_{3,1}t,g_{4,0}t^2,g_{5,1}t^3,\ldots, g_{2N,0}\,t^{2N-2}\}$, of all coefficients multiplied by the appropriate powers of $t$ to give them all the same dimension, and organised in growing order. Then we can write the right-hand side of \eq{a0appe} as $\vec a_0 \cdot \vec g$, the improved arc as $\vec a_{\text{imp}} \cdot \vec g$ (with $\vec a_{\text{imp}}=\{1,-1,0,0,0,\cdots\}$), and \eq{eq:darc2} as $\vec c \cdot A \cdot \vec g$, with $\vec c$ a vector collecting the $c_{n,k}$ coefficients and $A$  a  square matrix of size $(N^2+5N)/2$, defined by \eq{eq:darc2}. With this definition the improvement formula becomes,
\begin{equation}\label{eq:matrix}
  \vec a_0\cdot \vec g  + \vec c\cdot A\cdot \vec g = \vec a_{\text{imp}}\cdot \vec g\quad\quad \Rightarrow 
    \quad\quad \vec c  = (\vec a_{\text{imp}} - \vec a_0).A^{-1}\,,
\end{equation}
since the formula must be valid for all values of $\vec g$. This algorithm can be implemented for different $N$.

To find null constraints, the procedure is very similar, but with $\vec a_{\text{imp}}\to \vec 0$, since all coefficients must be cancelled. For instance, in the case with no gravity, we could cancel order by order all coefficients in $\partial^2_k a_0$ using higher arcs. The associated coefficients $\tilde c_{n,k}$ are $\vec{\tilde c} =\overrightarrow{\partial^2_k a_0}\cdot A^{-1}$, where the $A$ matrix is the same as above, while $\overrightarrow{\partial^2_k a_0}$ is the vector of coefficients of $\vec g$ in $\partial^2_k a_0(s,t)$.

In this work we do not use these null constraints since, as described below \eq{eq:onlyafew}, smearing the improved arc with high Legendre polynomials gives zero on thee IR side and something non-trivial on the UV side, thus automatically delivering null UV conditions.

\section{Bounds from smearing}\label{app:smerding}

In this appendix we provide a review on the semi-definite optimisation technique we used, the differences from the approach of Ref.~\cite{Caron-Huot:2021rmr} and further details on the numerical procedure.

\subsection{Smearing and semi-definite problem}\label{app:smerding1}
 A first difference w.r.t.\ Ref.~\cite{Caron-Huot:2021rmr} is that we use  polynomials in $t$ rather than  $p$, with $t=-p^2$. 
 In terms of building functionals, both approaches are equivalent -- we opted for $t$ to preserve the property described below \eq{eq:onlyafew}.
The constraints in impact parameter space described in Ref.~\cite{Caron-Huot:2021rmr} can still be added to our procedure, but we found that they have little impact on our bounds.

To define a semi-definite optimisation problem we start by writing $f(t)$ as in Eq.~(\ref{eq:measure}).
We can explicitly integrate the single functions both in the UV as in Eq.~(\ref{eq:UVintsdone}), defining the  vectors $\vec W$ and $\vec V$ in the space spanned by the $P_j$'s
\begin{equation} W_{j,\;\ell}(s') =
     \int_{-t_{\text{max}}}^0 \dd t  \;\left(\frac{t}{t_{\text{max}}}\right)\,  P_{j}\left(\frac{t}{t_{\text{max}}}\right) I_{\ell}^{N}(s',t),
\end{equation} 
and in the IR 
\begin{equation} 
    V_j =  \int_{-t_{\text{max}}}^0 \dd t\, \left(\frac{t}{t_{\text{max}}}\right) \;P_{j}\left(\frac{t}{t_{\text{max}}}\right)\, a_{0}^{\text{imp},N}(t) 
\end{equation} 
such that the
positivity requirement is written as 
\begin{equation}\label{eq44app}
    \sum_{j=0}^{j_{\rm{max}}} b_j W_{j,\;\ell}(s') \geq 0, \quad \forall \,s',
    \; \ell, \; \implies  \sum_{j=0}^{j_{\rm{max}}} b_j V_i \geq 0.
\end{equation} 
Having complete freedom on the choice of $f(t)$ and
therefore $b_j$, we can try to find a function which makes the bound
evident.

This is a semi-definite optimisation problem, where we have a target
vector $V_j$ and a vector we want to optimise $b_j$ to get $b_j V_j
\geq 0$, subject to some constraints $W_{j,\;\ell}(s') \succcurlyeq0$. 

Writing the typical
form of $V_j$ better reveals the details of the numerical procedure involved,
because the IR arc contains information about $g_2$, $g_3$ and $g_4$ for the 2D bound, depending on which coefficient we want to target. For example,
\begin{equation} \label{appCtarget}
    V_j =   \int_{-t_{\text{max}}}^0 \dd t\,\left(\frac{t}{t_{\text{max}}}\right) \;P_{j}\left(\frac{t}{t_{\text{max}}}\right)\,  \left( 2
    g_2 - g_3 t + 8 g_4 t^2\right)\,, 
\end{equation} 
and we can ask to optimise for
example the value of $g_3/g_2$ for a fixed value of $g_4/g_2$.\\ Furthermore the
boundary of the allowed region is found when the inequality is saturated.
Therefore we can implement an additional bisection algorithm, where we choose a
value of $g_4/g_2$ and multiple values of $g_3/g_2$. If the optimisation problem
results in a positive value for $b_j V_j$ then the value is allowed,
otherwise it is prohibited.

\begin{figure}
    \centering
    \includegraphics[width=0.7\textwidth]{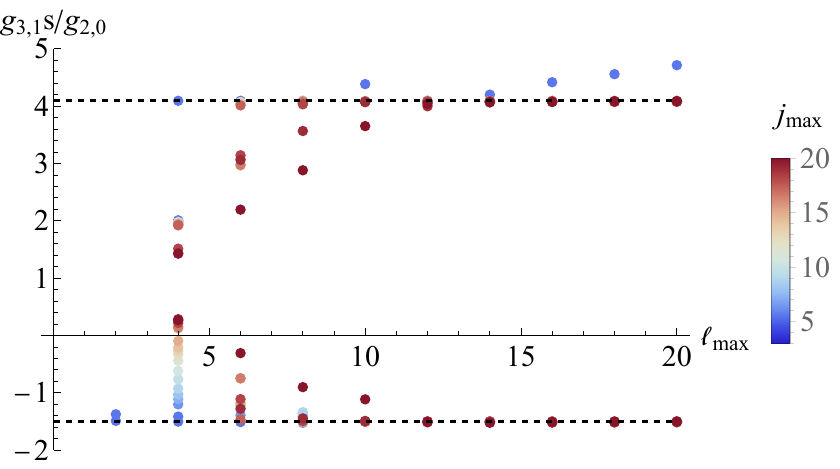}
    \caption{\it \footnotesize Convergence of the 1D bound for $g_{3,1}s/g_{2,0}$ for various values of $\ell_{\rm max}$. The color bar refers to the number of elements of the polynomial basis used, $j_{\rm max}$. 
   For fixed $\ell_{\rm max}$, a large enough basis is needed to obtain a better bound. Vice versa, for small $\ell_{\rm max}$ the bound is not conservative, since not enough conditions have been imposed in the UV. 
   For large enough values of both $\ell_{\rm max}$ and $j_{\rm max}$, bounds converge to the correct value (dashed black line).}\label{fig:convergence}
\end{figure}

In Figure~\ref{fig:convergence} of this appendix we show how the $g_{3,1}s/g_{2,0}$ bounds  
behave as functions of the algorithm parameters $\ell_{\rm max}$ and $j_{\rm max}$. For small $\ell_{\rm max}$ and a large basis of polynomials $j_{\rm max}$, the bounds might be too strong (non conservative), since not enough conditions have been imposed. For larger values of both parameters,  they asymptotically approach the correct values (dashed black lines).

\subsection{Implementation of the numerical procedure} 
To set up the problem numerically and ensure positivity for every $s'$ we performed an expansion in $s'$ as described in Sec.~\ref{sec:boundswc}. Here we provide additional details.

In the UV, we can change variables 
to $x=t/t_{\text{max}}$,
\begin{equation}
    W_{j,\;\ell}(s') =
    t_{\text{max}} \int_{-1}^0 \dd x\; x \;  P_{j}(x) I_{\ell}^{N}(s',(x \,t_{\text{max}})).
\end{equation}
Inspecting the function $I_{\ell}^N$ reveals that it could be written as an overall dimensionful $1/s'^k$ times a function of the adimensional variable $ x\, t_{\text{max}}/s'$.
Being $t_{\text{max}}/s' \ll 1$ and $x\in [-1,0]$, we can expand with a Taylor series the function in the integral, giving
 \begin{equation}
     s'^{k}I_{\ell}^{N}(s',(x \,t_{\text{max}})) = g \left(\frac{x\, t_{\text{max}}}{s'}\right)=\sum_{j=0}^M a_{j} \left(\frac{x\, t_{\text{max}}}{s'}\right)^j +  \left(\partial_{x}^{M+1}g\Big|_{t=\zeta}\right) \left(\frac{\zeta\, t_{\text{max}}}{s'}\right)^{(M+1)},
 \end{equation}
where the second piece is the rest of the Taylor expansion and gives us an estimate of the error.
Therefore we can expand the whole partially improved arc $I_{\ell}^{N}$ as a polynomial in $t_{\rm max}/s'$, up to a certain order $M$.

The parameter $M$ is chosen such that $M\approx 2N$, i.e.\ we require consistency in the size of the terms we omit, since
the improvement omits terms of order $t_\text{max}^{2N+2}$, and the Taylor expansion omits terms of order $t_\text{max}^{M+1}$.
For all our bounds, we consistently discard higher powers of $t_\text{max}$ if the leftover terms 
in the improvement and the Taylor expansion are of order~$10^{-9}$. Table~\ref{tab:consistentparams} shows some of the various consistent choices we make.

\begin{table}[h]
    \centering
    \begin{tabular}{c c c c c }
        $t_\text{max}$ & $N$ & $M$ & $(t_\text{max}/s)^M$ & first discarded term in the improvement\\
        \hline
        $0.5t_*$ & $6$ & $12$ & $3\times10^{-9}$ & $2\times10^{-8}g_{14,0}$\\
        $0.8t_*$ & $8$ & $16$ & $8\times10^{-9}$ & $7\times10^{-8}g_{18,0}$\\
        $0.99t_*$ & $12$ & $20$ & $5\times10^{-9}$ & $1.6\times10^{-9}g_{26,0}$\\
    \end{tabular}
    \caption{\it \footnotesize Consistent parameter choices for the partial improvement and Taylor expansion
    of the arcs, such that the discarded pieces are of similar order. These are the choices we use
    throughout the article.}\label{tab:consistentparams}
\end{table}

This expansion in $t_\text{max}/s'$ serves two purposes. First it speeds up the numerical calculation because now $W_{j,\ell}$ after the expansion is written as a polynomial in $s'$ with a global pre-factor $1/s'^{M+1}$ and can be easily integrated analytically.
Second it allows us to require positivity for a generic polynomial in $s'$, for $s'\in [1, \infty]$. Without the expansion we require to optimise a generic function for all values of $s'$ and the only way is by taking a discrete grid of values for it. This discretisation not only makes the numerical problem harder but it also weakens the bound, as argued in Ref.~\cite{Caron-Huot:2021rmr}.

Given that we require positivity for each
\begin{equation}
    \sum_{j=0}^{j_{\rm{max}}} b_j W_{j,\;\ell}(s')\geq 0
\end{equation}
we can solve now the equivalent problem expanded in $t_{\rm max}$
\begin{equation}
    s'^{M+1} \sum_{j=0}^{j_{\rm{max}}} b_j W_{j,\;\ell}(s')\geq 0.
\end{equation}

To obtain the various
bounds found in this work we set up the problem as above and we use the numerical optimisation algorithm  {\tt sdpb}~\cite{Simmons-Duffin:2015qma,Landry:2019qug}.\\
We used the constraint as input  for multiple values of $\ell$, optimising a polynomial in $s'$.
We built a matrix with rows made from $1\times1$ matrices for the discrete values of $\ell$,
and a column for each $j$ of the basis chosen for $f(t)$. 
While setting up the problem we are free to choose a normalisation for the target vector~$b_j$. We
choose to always take the following normalisation condition,
\begin{equation}
    \sum_{j=0}^{j_{\rm{max}}}b_j \int_{-t_{\rm max}}^0 \dd t \;  \left(\frac{t}{t_{\rm max}}\right)\, P_{j}\left(\frac{t}{t_{\rm max}}\right) \; t = \pm1, 
\end{equation} 
which corresponds to the ``vector'' associated with
$g_3$, giving for example the bound with 
the $+$ sign in the normalisation,
\begin{align} 
    &2\left( \sum_{j=0}^{j_{\rm{max}}}b_j\int_{-t_{\rm max}}^0 \dd t \;  \left(\frac{t}{t_{\rm max}}\right)   P_{j}\left(\frac{t}{t_{\rm max}}\right)\right)  +\\
    &+8\frac{g_4}{g_2} \left( \sum_{j=0}^{j_{\rm{max}}}b_j \int_{-t_{\rm max}}^0 \dd t \;  \left(\frac{t}{t_{\rm max}}\right)\;  P_{j}\left(\frac{t}{t_{\rm max}}\right) \;  t^2\; \right)  \geq \frac{g_3}{g_2}, 
\end{align} 
or, choosing the
other normalisation, a lower bound.\\
For the bounds with gravity, we find it useful to restrict the parameters $b_j$ to be at most of order $10^6$ in absolute value. This facilitates numerical convergence and avoids fake solutions. We implement this in {\tt sdpb} by adding rows and columns to the positive matrix for each $j$, forcing $\pm b_j\geq -10^6$.\\
We include typical values used in the Table~\ref{tb:values}. The algorithm for {\tt sdpb} taken from Ref.~\cite{Simmons-Duffin:2015qma,Landry:2019qug} is set up with the default parameters except the maximum number of iterations that has been increased for the case with gravity from $500$ to $1000$.
\begin{table}[h] 
\begin{center} \begin{tabular}{ c c c c c c}
 & Fig.~\ref{fig:tmaxdependence}& Fig.~\ref{fig:g3g4} & Fig.~\ref{fig:gravd6} & Fig.~\ref{fig:gravd562}\\ 
\hline $j_{\rm max}$ & 7 &  10 & 4, 6, 8 & 8 \\ 
$ \ell_{\text{max}}$ & 14 & 100 & 400 & 400 \\
$N$   & 3, 8  &  6 & 6, 8, 12 &12\\
\end{tabular}
\end{center} 
\caption{Values of the parameters of the improvement algorithm for all the plots in this article. Refer to Table~\ref{tab:consistentparams} for the choices of $N$, $M$, and $t_\text{max}$.}\label{tb:values}
\end{table}

\section{Polynomial improvement}\label{sec:apppol}
In this section we propose a radically different approach to the improvement at finite~$t$, based on  summing arcs weighed by polynomials, as in \eq{eq:geralimpPOL}, in such a way that the $t$-integral is improved, see \eq{imrpvementpolynomial}.
Similarly, a different set of polynomials would give $\alpha {\kappa}+ \beta
    g_{2,0}+\gamma g_{3,1}+g_{4,0}\delta$ in the IR, or any other linear combination of finitely many coefficients.

Differently from what is discussed in the main text, here the improvement takes place at the level of the integrated dispersion
relations rather than as a function.   

Such polynomials can be built at any finite order of partial improvement $N$. We describe this in more detail below. It is not obvious, however, that such polynomials survive the  $N\to \infty$ limit, as we now show with an example.  
Indeed, consider a polynomial 
$p^s(t)$, such that  $\int_{-t_{\text{max}}}^0 \dd t\, p(t) a_0(s,t)$ equals the right-hand
side of \eq{imrpvementpolynomial}. In some sense, $p^s(t)$ is a polynomial that improves the arc, without relying on full crossing symmetry. Such a polynomial would have to be orthogonal
(w.r.t.\ the scalar product $\langle f,g\rangle\equiv \int_{-t_{\text{max}}}^0 f(t) g(t)$) to all
monomials $t^2,t^3,t^4,\cdots,t^{N}$. As explained in Section~\ref{subsec:MS}, the
M\"untz-Sz\'asz theorem then implies that  as we take $N\to \infty$, then $p^s(t)\to 0$, since the monomials
$t^2,t^3,t^4,\cdots$ are a basis. In conclusion, such a polynomial does not exist.

Instead, a non-trivial version of \eq{imrpvementpolynomial} can be obtained using full crossing symmetry. The polynomials $p_n(t)$ must have at least $n+1$
independent terms -- the same as the number of derivatives needed to
improve the arcs via \eq{eq:geralimp}, which follows simply from our counting of
parameters introduced by each new arc. 
To find these polynomials, we insert \eq{eq:geralimpPOL} evaluated in the IR into
\eq{imrpvementpolynomial} and demand that the left-hand and right-hand sides match for all values of the coefficients $g_{n,q}$, up to a given order $n<N$. 
For the first two polynomials, for instance,
\begin{equation}
    p_0(t)=\alpha_{0}^0+\alpha_{0}^1 t\,\quad 
    p_1(t)=\alpha_{1}^0+\alpha_{1}^1 t+\alpha_{1}^2 t^2\,,
\end{equation}
we find,
\begin{eqnarray}
    \int_{-t_\text{max}}^0\sum p_n(t)a_n(t)&=&g_{2,0}\left(\alpha_0^0 t_\text{max}+\alpha_0^1 \frac{t_\text{max}^2}{2}\right)
    +g_{3,1}\left(\alpha_0^0 \frac{t_\text{max}^2}{2}+\alpha_0^1 \frac{t_\text{max}^3}{3}\right)\\
   && +g_{4,0}\left(\alpha_0^0 \frac{2t_\text{max}^3}{3}+\alpha_0^1 \frac{t_\text{max}^4}{2}+\alpha_1^0 t_\text{max}+ \alpha_1^1 \frac{t_\text{max}^2}{2}+\alpha_1^2 \frac{t_\text{max}^3}{3}\right)
   +\cdots\nn
\end{eqnarray}
Now, demanding that this reproduces \eq{imrpvementpolynomial} (with $\kappa=0$), leads to 
\begin{eqnarray}
    p_0(t)&=&-\frac{t}{t_{\rm max}} \left(\frac{6 \beta}{t_{\rm max}}-\frac{12 \gamma}{t_{\rm max}^2}\right)-\frac{4 \beta}{t_{\rm max}}+\frac{6 \gamma}{t_{\rm max}^2}\, ,\\
p_1(t)  &=&  t^2 \left(\frac{26 \beta}{t_{\rm max}}-\frac{102 \gamma}{t_{\rm max}^2}\right)-t \left(\frac{504 \gamma}{5 t_{\rm max}}-\frac{132 \beta}{5}\right)+\frac{21 \beta t_{\rm max}}{5}-\frac{72 \gamma}{5}\,.
\end{eqnarray}
This gives $ a^\text{imp-pol}_0=\beta g_{2,0}+\gamma g_{3,1}
+g_{6,0}(3/5\beta t_{\rm max}^4-\frac{12}{5} \gamma t_{\rm max}^3 )
+\dots$, where the dots indicate more irrelevant operators, and no dependence on $g_{4,0}$ is left (in fact these polynomials also cancel any dependence in $g_{5,1}$ and $g_{6,2}$).
Repeating the same procedure, including higher arcs and other polynomials, one finds an algorithm  to compute all polynomials $p_n(t)$.  If such formula could be re-summed into the UV side it would provide a formula improved at all orders which could be readily be used, similarly to that of Ref.~\cite{Caron-Huot:2022jli}, but completely at finite $t$: a full rather than a partial improvement.

\newpage
\bibliographystyle{JHEP} \bibliography{bibs}  

\end{document}